\documentclass[aps,preprint]{revtex4}
\usepackage[colorlinks,linkcolor=black,citecolor=blue]{hyperref}
\usepackage{amsfonts}
\usepackage{amsmath}
\usepackage{amssymb}
\makeatletter

\newcommand{\Rmnum}[1]{\expandafter\@slowromancap\romannumeral #1@}
\makeatother
\usepackage{graphicx}
\usepackage{grffile}
\usepackage{epstopdf}
\usepackage{float}
\usepackage{ulem}
\usepackage{cancel}
\usepackage{xcolor}
\usepackage{cleveref}

\usepackage{bbold}
\usepackage{bm}
\usepackage{subfigure}

\begin{document}

	\title{Thermodynamics and Microstructures of Euler-Heisenberg Black Hole in a Cavity}
	\author{Qin Yu}
	\email{yq@stu.scu.edu.cn}
	\author{Qi Xu}
	\email{xuqi@stu.scu.edu.cn}
	\author{Jun Tao}
	\email{taojun@scu.edu.cn}
	\affiliation{Center for Theoretical Physics, College of Physics, Sichuan University, Chengdu, 610065, China}

\begin{abstract}
The Euler-Heisenberg black holes with quantum electrodynamics (QED) correction are embraced by a cavity in this paper, which serves as a boundary of the black hole spacetime and contributes to the equilibrium of the system. We explore the thermodynamic properties of the black hole, including the phase transitions and phase structures. The small/large black hole phase transition occurs for a negative QED parameter, while the reentrant phase transition can be observed for a small positive QED parameter. Then the thermodynamic geometry is investigated to diagnose microscopic interactions of black hole thermodynamic systems. For the reentrant phase transition, the small black holes are dominated by repulsion for the first-order coexistence curve, while the interaction between the small black hole molecules could be attractive or repulsive for the small/large black hole phase transition.
\end{abstract}
\maketitle

\section{Introduction}
The investigation of black hole thermodynamics has shown crucial importance to quantum gravity and general relativity, based on the pioneering works by Hawking, Bekenstein and others \cite{Hawking1971,Bekenstein1972,Hawking1974,Bardeen1973}. Between the thermal anti-de Sitter (AdS) space and black holes, the Hawking-Page phase transition was observed in Schwarzschild-AdS black holes for the first time  \cite{Hawking1983}. Inspired by the AdS/CFT correspondence between gravity systems and the conformal field theory \cite{Maldacena1998,Gubser1998,Witten1998}, the Hawking-Page phase transition can be interpreted as a confinement/deconfinement transition of gauge field \cite{Witten1998a}. Thermodynamic properties and phase structures of black holes were deeply researched then, which has gained significant attention in recent years.

The relation between black hole thermodynamics and conformal field theory needs further discussion. For a fixed temperature at infinity, black holes are thermodynamically unstable in asymptotically flat space \cite{Wang2020a}. Enclosing black holes with a cavity is a useful way to give solutions of it. Considering a heat bath around the cavity, the surface gives a fixed temperature. Black holes can be thermodynamically stable then, as in the case of AdS space, for that the horizon of the black hole can be near the fixed temperature point \cite{Lundgren2008}. This method is useful to investigate the holography in asymptotically flat and AdS space as well. The Schwarzschild black hole in a cavity was researched by York for the first time \cite{York1986}. And comparing with the Schwarzschild-AdS black holes, there are overall similarity between the phase structures of these two cases. The Reissner-Nordstr$\ddot{\text{o}}$m (RN) black holes in a cavity showed a Hawking-Page phase transition in the grand canonical ensemble \cite{Braden1990} and a van der Waals phase transition in the canonical ensemble \cite{Carlip2003}, and it is closely similar to the phase structures of a RN-AdS black hole. There is almost no difference in phase structures between Gauss-Bonnet black holes in a cavity and Gauss-Bonnet AdS black holes as well \cite{Wang2020a,He2017}. However, some differences exist, such as phase structures of Born-Infeld black holes \cite{Wang2019,Wang2019a,Liang2020}. For example, a large/small/large black hole reentrant phase transition \cite{Astefanesei2022,Altamirano2013,Bai2022a} could be observed for Born-Infeld AdS black holes, while it can't occur for any Born-Infeld black hole in a cavity. Between black holes in a cavity and with the AdS boundary, the similarity or difference of phase structures may have interesting implications for black hole thermodynamics and holography \cite{Wang2021}. These investigations have aroused our interest to explore black holes in a cavity.

In addition, the statistical description of the black hole microstates is still not entirely clear. Following the pioneering work by Weinhold \cite{Weinhold1975}, to represent the thermodynamic fluctuation theory, Ruppeiner constructed a Riemannian thermodynamic entropy metric and developed a systematic method to calculate the Ricci curvature scalar $R$ of the metric \cite{Ruppeiner1995}. The sign of Ruppeiner invariant is relative to the intermolecular interaction: $R>0$ denotes repulsion while $R<0$ denotes attraction, and $R=0$ indicates no interaction \cite{Ruppeiner2010}. Whereafter, the Ruppeiner approach was applied in various black holes \cite{Aman2006,Sarkar2006,Biswas2010,Banerjee2011,Niu2012,Wei2013,Suresh2014,Mansoori2014,Wei:2015iwa,Chaturvedi2017,KordZangeneh2018,Wei:2019uqg,Wang2020,Wang2022,Huang2022}. By employing the Ruppeiner geometry, small black hole behaves like a bose gas with reentrant phase transitions. However, it behaves like a quantum anyon gas \cite{Mirza2009} with usual first-order phase transitions, so the phase transition may be affected by the microstructure of black holes \cite{KordZangeneh2018}. These studies show good examples combining the phase transition with the microstructure of black holes. 

To understand the thermodynamics and microstructure of black holes, we pay close attention to the nonlinear electrodynamics for its extensive research. After first introduced by Born and Infeld \cite{Born1934}, nonlinear electrodynamics can be concerned in the study of black holes coupling with general relativity. The Euler-Heisenberg electrodynamics is a particular example of nonlinear electrodynamics as well \cite{Heisenberg1936}. Then Schwinger reformulated the modified Lagrange function for constant fields within the quantum electrodynamics (QED) framework \cite{Schwinger1951}. It is of particular interest to combine the Euler-Heisenberg electrodynamics with the study of black holes, such as the asymptotically flat black holes in Euler-Heisenberg electrodynamics \cite{Yajima2001}. Recently, the Euler-Heisenberg black holes in AdS space with QED effects was considered to investigate the influence on black hole phase transition and Ruppeiner geometry in the extended phase space \cite{Ye2022}. Some other relevant works has been considered in the literature \cite{Dai2022,Stefanov2007,Ruffini2013,Magos2020,Li2022,Maceda2019,Guerrero2020,Maceda2021,Breton2021,Nashed2021}. Inspired by these researches, we investigate the phase structures of Euler-Heisenberg black holes in a cavity and find some interesting phenomenons, especially the reentrant phase transition. Furthermore, from a microscopic point of view, the interaction is repulsive between the small black hole molecules in the reentrant phase transition case.

The structure of this paper is as follows. In Sec. \ref{sec:2}, we derive the thermodynamics of Euler-Heisenberg black holes in a cavity. In Sec. \ref{sec:3}, black hole phase transitions about the QED parameter and charge are analyzed. Then different phase structures and phase diagrams are investigated for positive and negative QED parameter, respectively. In Sec. \ref{sec:4}, the microstructure of the black hole is studied by Ruppeiner geometry. Finally, we summarize the findings and make some discussions in Sec. \ref{sec:Discussion-and-Conclusions}. Here we set $G=\hbar =c=1$ for simplicity in this paper.

\section{Thermodynamics}
\label{sec:2} 
The metric of the static spherically symmetric charged Euler-Heisenberg black hole can be written as \cite{Magos2020} 
\begin{equation}
	\mathrm{d}s^2=-f(r){\mathrm{d}t}^2+\frac{{\mathrm{d}r}^2}{f(r)}+ r^2{\mathrm{d}\theta }^2+ r^2 \sin ^2\theta{\mathrm{d}\phi }^2,
\end{equation}
where
\begin{equation}
	f(r)=1-\frac{2M}{r}+\frac{q^2}{r^2}-\frac{aq^4}{20r^6}.
\end{equation}
The parameter $M$ is the ADM mass which can be obtained by solving the equation $f(r_h)=0$, while $r_h$ is the event horizon of the black hole. The parameter $q$ is the electric charge and $a$ represents the strength of the QED correction \cite{Ruffini2013}. The metric reduces to the RN black hole case when the QED parameter $a$ is equal to zero.

Following the derivations of Euclidean action for a NLED charged black hole inside a cavity in Ref. \cite{Wang2019}, the thermal energy of the Euler-Heisenberg black hole in a cavity with the radius $r_B$ takes the form as
\begin{align}
	E=r_B\left[1-\sqrt{f(r_B)}\right].
\end{align}
We can obtain the ADM mass by extending the cavity radius to infinity,
\begin{align}
	\lim_{r_B \to \infty} E=M.
\end{align}

The temperature of these black hole systems can be written as \cite{Wang2019a} 
\begin{align}
	T=T_h/\sqrt{f(r_B)},\label{Temperature}
\end{align}
and the Hawking temperature $T_h$ of Euler-Heisenberg black holes is
\begin{align}
T_h=\frac{1}{4 \pi r_h}\left(1-\frac{q^2}{{r_h}^2}+\frac{a q^4}{4 {r_h}^6}\right).
\end{align} 
Then, it is natural to establish the first law of thermodynamics
\begin{align}
	\mathrm{d}E=T\mathrm{d}S+\Phi \mathrm{d}q,
\end{align}
where the entropy $S$ is expressed as $S=\pi r_h^2$. The electric potential can be determined by differentiating the thermal energy with respect to the charge, which can be expressed as
\begin{align}
	\Phi=\left(\frac{\partial E}{\partial q}\right)=\frac{r_B}{2\sqrt{f(r_B)}}\left[-\frac{aq^3(r_B^5-r_h^5)}{5r_B^6r_h^5}+\frac{2q(r_B-r_h)}{r_B^2r_h^2}\right].
\end{align}

The investigation of black hole phase transitions usually requires the analysis of critical points, which can be determined by the following equations
\begin{equation}
	\left(\frac{\partial T}{\partial r_h}\right)_{r_B,a,q}=0,\quad\left(\frac{\partial^2T}{\partial r_h^2}\right)_{r_B,a,q}=0.\label{critical point}
\end{equation}
However, the analytical solution of critical points is difficult to be derived from the temperature of the Euler-Heisenberg black hole in a cavity, which impels us to solve it numerically. The type of phase structure can be roughly determined by the number of critical points, and the critical point can be combined with free energy for analysis.
The free energy of black holes takes the form as
\begin{align}
	F=E-TS.
\end{align}
Several phases exist if there is more than one value of the free energy for a given temperature. Among these phases, the phase with the lowest free energy is globally stable. Then phase transitions can be judged by investigating whether there are multiple branches with the lowest free energy. At the critical point, the first derivative of the free energy curve is continuous, and the second derivative is discontinuous.

In following sections, we will further study the phase transition of Euler-Heisenberg black hole in a cavity for different QED parameters. The microstructure of this black hole will be discussed by the Ruppeiner geometry as well.

\section{Phase transition and phase structure}\label{sec:3}

In this section, we shall study the phase transition of Euler-Heisenberg black hole in a cavity. The phase behaviors of black hole thermodynamic systems will be discussed for different QED parameters. 

\begin{figure}[htbp]
	\centering
	\subfigure[]{\begin{minipage}{8cm}
		\includegraphics[width=0.9\linewidth]{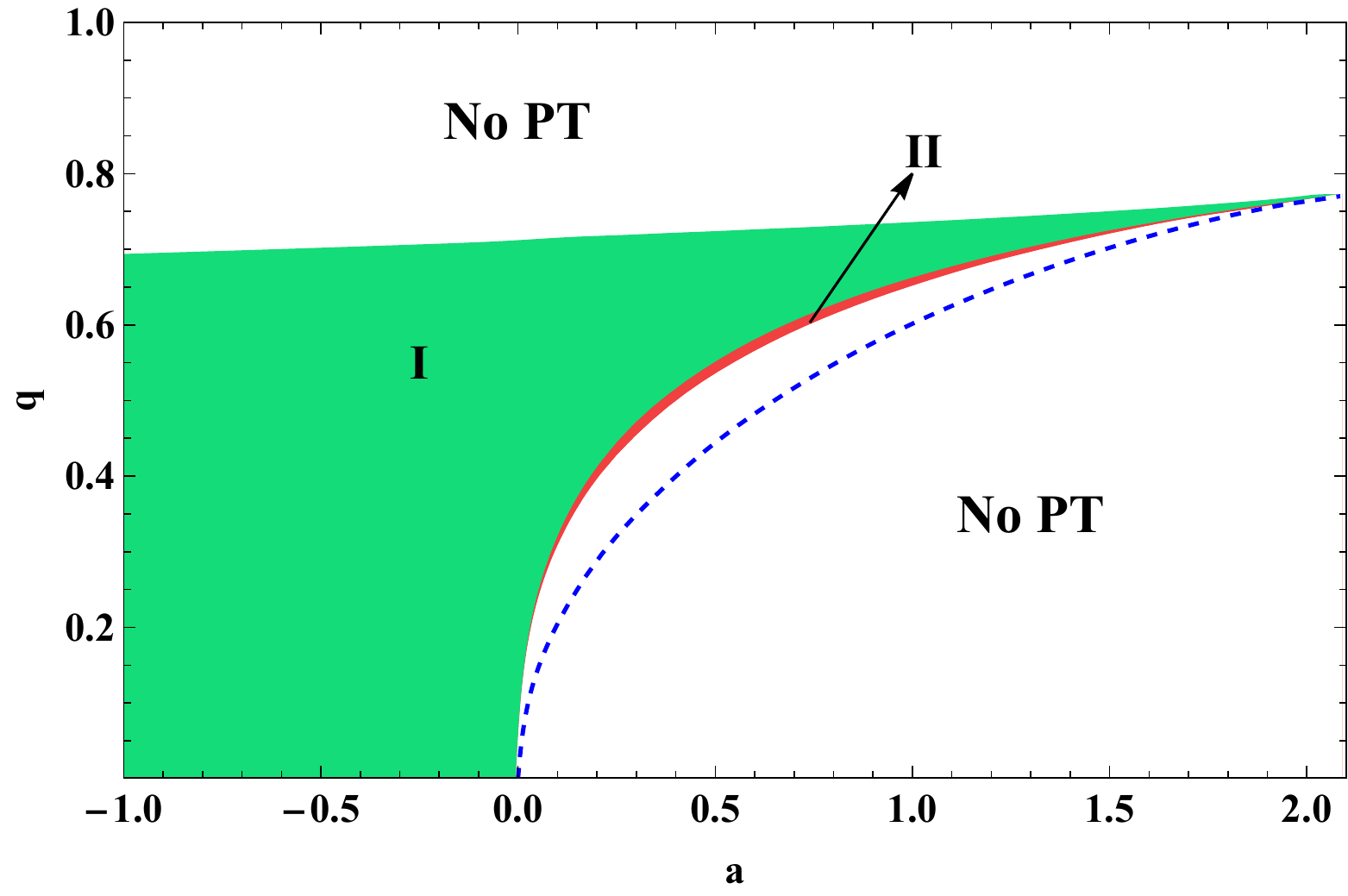}\label{qvaa}
	\end{minipage}}
	\subfigure[]{\begin{minipage}{8cm}
		\includegraphics[width=0.9\linewidth]{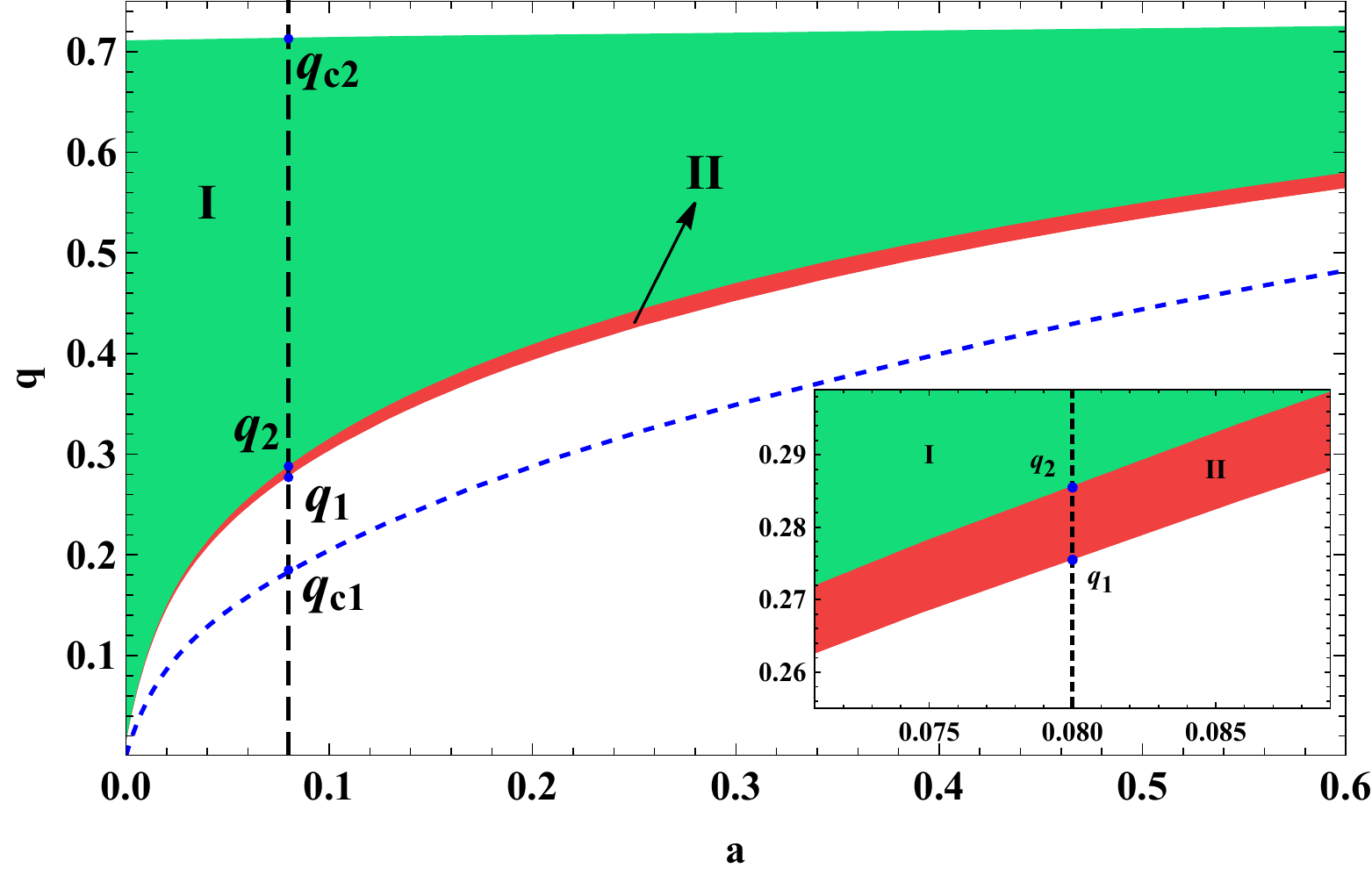}\label{qvab}
	\end{minipage}}
	\caption{Regions of different phase transitions for $q$ and $a$. \textbf{Left Panel (a):} The white region represents the system without phase transition (PT). The green region \Rmnum{1} corresponds to the case that a single first-order phase transition occurs. The reentrant phase transition happens in the red region \Rmnum{2}. The blue dashed line corresponds to the unstable critical point for $a>0$, which does not present the occurrence of phase transitions. \textbf{Right Panel (b):} The black dashed line $a=0.08$ intersects the blue dashed line at $q_{c1}$, and intersects the boundary of regions at $q_1$, $q_2$, $q_{c2}$, respectively. The cavity radius is set to be $r_B=3$.}\label{qva}
\end{figure}

Based on features of the free energy, we find that phase transitions depend on the QED parameter $a$ and charge $q$ with a fixed cavity radius. Setting $r_B=3$ without loss of generality, show the phase transition types for different $q$ and $a$ in  Fig. \ref{qva}. The white region in Fig. \ref{qvaa} corresponds to the system without phase transition. In the green region \Rmnum{1}, a single first-order phase transition occurs.  The reentrant phase transition exists for $a>0$, which is shown as the red region \Rmnum{2}. The black hole (BH) jumps from a stable large BH phase to a small one through a zeroth-order phase transition, and finally returns to the large BH phase through a first-order phase transition. The blue dashed line is given by the unstable critical point of charge for $a>0$, which is not related to the occurrence of phase transitions.

There is a maximum value of QED parameter for the existence of phase transitions, which can be labeled as $a_{max}$. It is similar to the result in the Euler-Heisenberg-AdS black holes, where $a_{max}=\frac{32}{7}q^2$ \cite{Ye2022}. For $a>a_{max}$, there is no real number solution from Eqs. (\ref{critical point}), so phase transitions cannot exist for any value of charge $q$. It is difficult to obtain the solution analytically for the Euler-Heisenberg black hole in a cavity, which induces us to solve it numerically. The maximum value of QED parameter allowing for phase transitions is determined by the cavity radius, and we can obtain $a_{max}=2.0848$ for $r_B=3$ here.

For a fixed QED parameter, the BH phase undergoes different phase transitions as the charge increases. For $a<0$, a first-order phase transition happens, where the small BH phase turns into a large one as the temperature increases. For $a>0$, we take $a=0.08$ as an example in Fig. \ref{qvab}, which will be discussed in Sec. \ref{sec:3.1} as well. The BH phase remains stable for a small charge $0<q<q_1$. When the value of the charge increases to the range of the red region $q_1<q<q_2$, the reentrant large/small/large BH phase transition occurs. After entering the green region, the reentrant phase transition disappears for $q_2<q<q_{c2}$, and the single first-order phase transition takes place. For $q>q_{c2}$, there is no phase transition and one can no longer distinguish black holes between the small and large branches.

\begin{figure}[htbp]
	\subfigure[$a<0$]{\begin{minipage}{8cm}
			\includegraphics[width=0.9\linewidth]{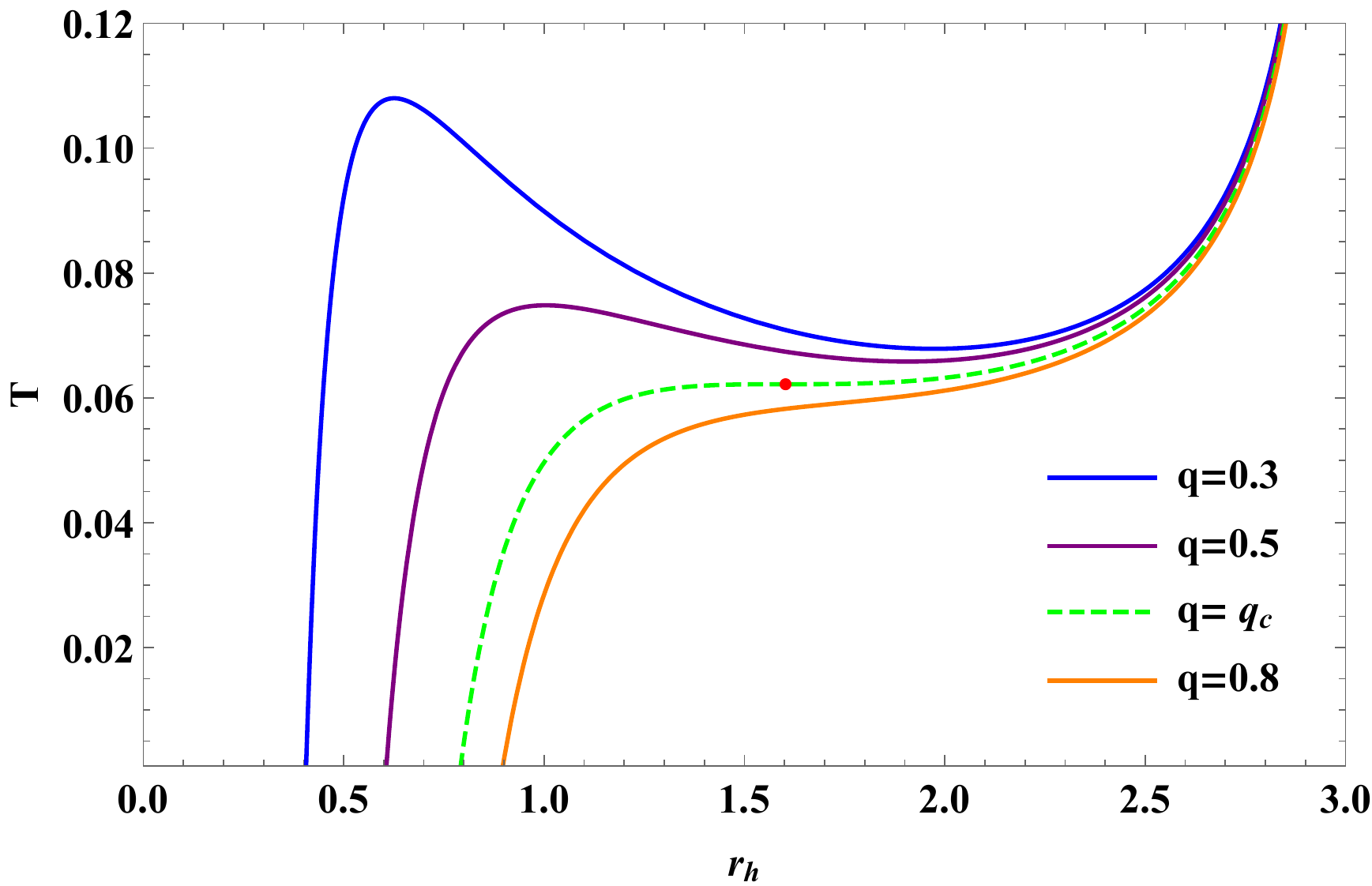}\label{Tvrha}
	\end{minipage}}
	\subfigure[$a>0$]{\begin{minipage}{8cm}
			\includegraphics[width=0.9\linewidth]{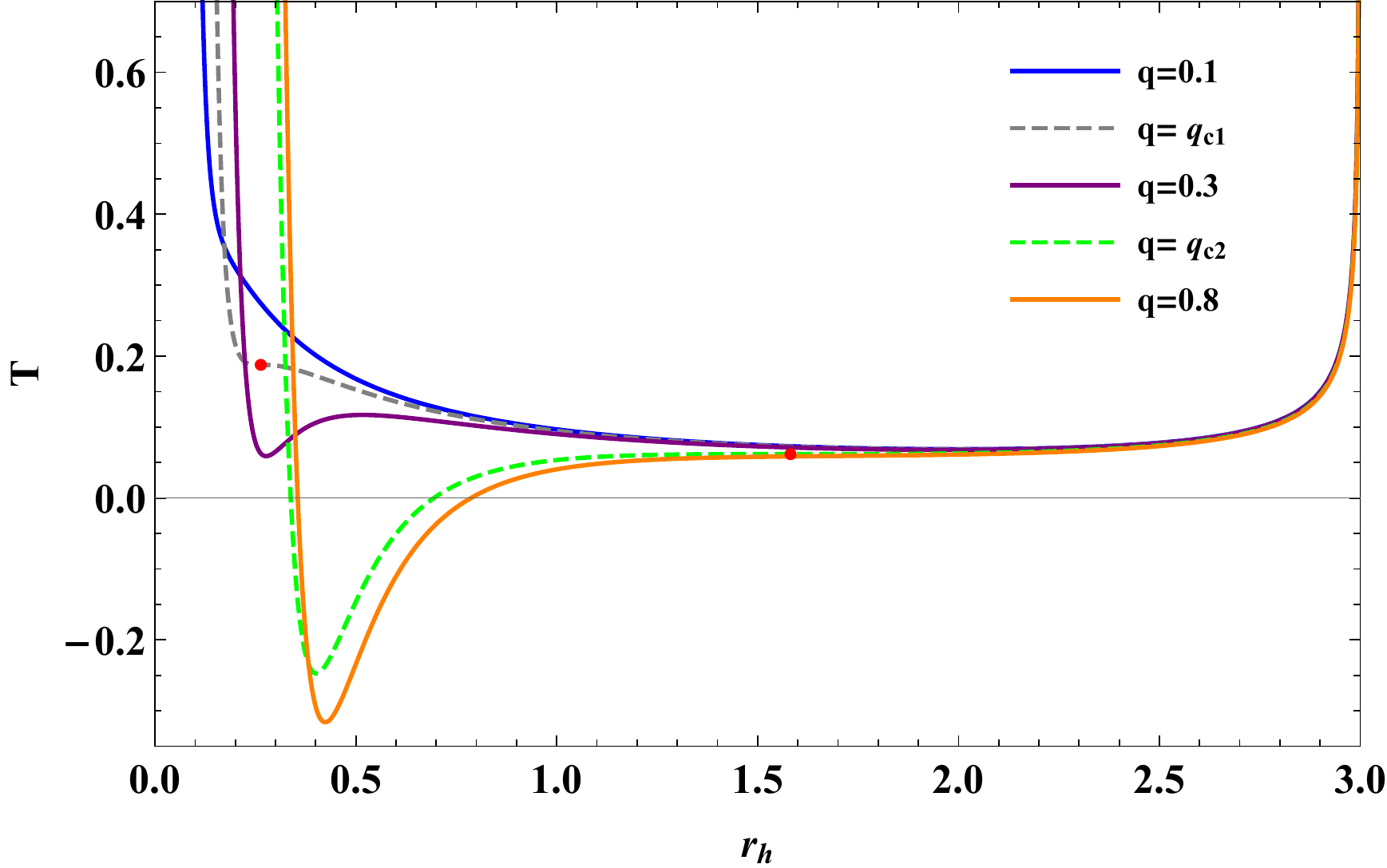}\label{Tvrhb}
	\end{minipage}}	
	\caption{Temperature curve of the Euler-Heisenberg black hole in a cavity versus horizon radius for different values of $q$. \textbf{Left Panel (a):} The parameter $a=-1$
	, and the charge $q$ is equal to $0.3,0.5,0.6909$ and $0.8$, respectively;  \textbf{Right Panel (b):} The parameter $a=0.08$, and the charge $q$ is equal to $0.1,0.1839,0.3,0.7098$ and $0.8$, respectively. The cavity radius $r_B$ is set to be $3$.}
	\label{Tvrh}
\end{figure}

Furthermore, the critical points of the Euler-Heisenberg black holes in a cavity can be investigated by analyzing the temperature curve. The temperature $T$ versus the horizon radius $r_h$ for negative and positive QED parameters is plotted in Fig. \ref{Tvrh}, where the cavity radius $r_B$ is set to be $3$. For $a=-1$, we obtain the solution of critical charge $q_c=0.6909$ from Eq. (\ref{critical point}); For $a=0.08$, we obtain two critical charges $q_{c1}=0.1839$ and $ q_{c2}=0.7098$. 

There are two extrema on the curve for $q<q_c$ as shown in Fig. \ref{Tvrha}. In this situation, one can divide the curve into three branches. The two sides of the curve correspond to the small BH (left side) and large BH (right side), respectively, while the middle branch represents the intermediate one. The heat capacity can be given as $C_q=T(\frac{\partial S}{\partial T})_q$, which indicates that the positive and negative slopes of temperature curve correspond to stable and unstable branches, respectively. Furthermore, two extrema will get closer as the charge increases, and then coincide at the critical point (red point). For $q>q_c$, there is no extremal point.

There are three extrema when $q_{c1}<q<q_{c2}$ in Fig. \ref{Tvrhb}, and two of them with larger $r_h$ will coincide at the critical point as the charge $q$ increases. In this case, there are four BH branches, including an unstable smaller branch, a stable small branch, a unstable intermediate branch and a stable large branch, which is a signal of possible phase transitions. For $q<q_{c1}$, there are one unstable BH branch and one stable branch, thus no phase transition occurs. There is no phase transition for $q>q_{c2}$ as well.

Following the discussions above, the Euler-Heisenberg black hole in a cavity undergoes different phase transitions for negative and positive QED parameters, and we will discuss these two cases respectively.

\subsection{$a>0$ case}
\label{sec:3.1}
\begin{figure}[htbp]
	\centering
	\subfigure[$q_{c1}<q<q_1$]{\begin{minipage}{8cm}
			\includegraphics[width=0.9\linewidth]{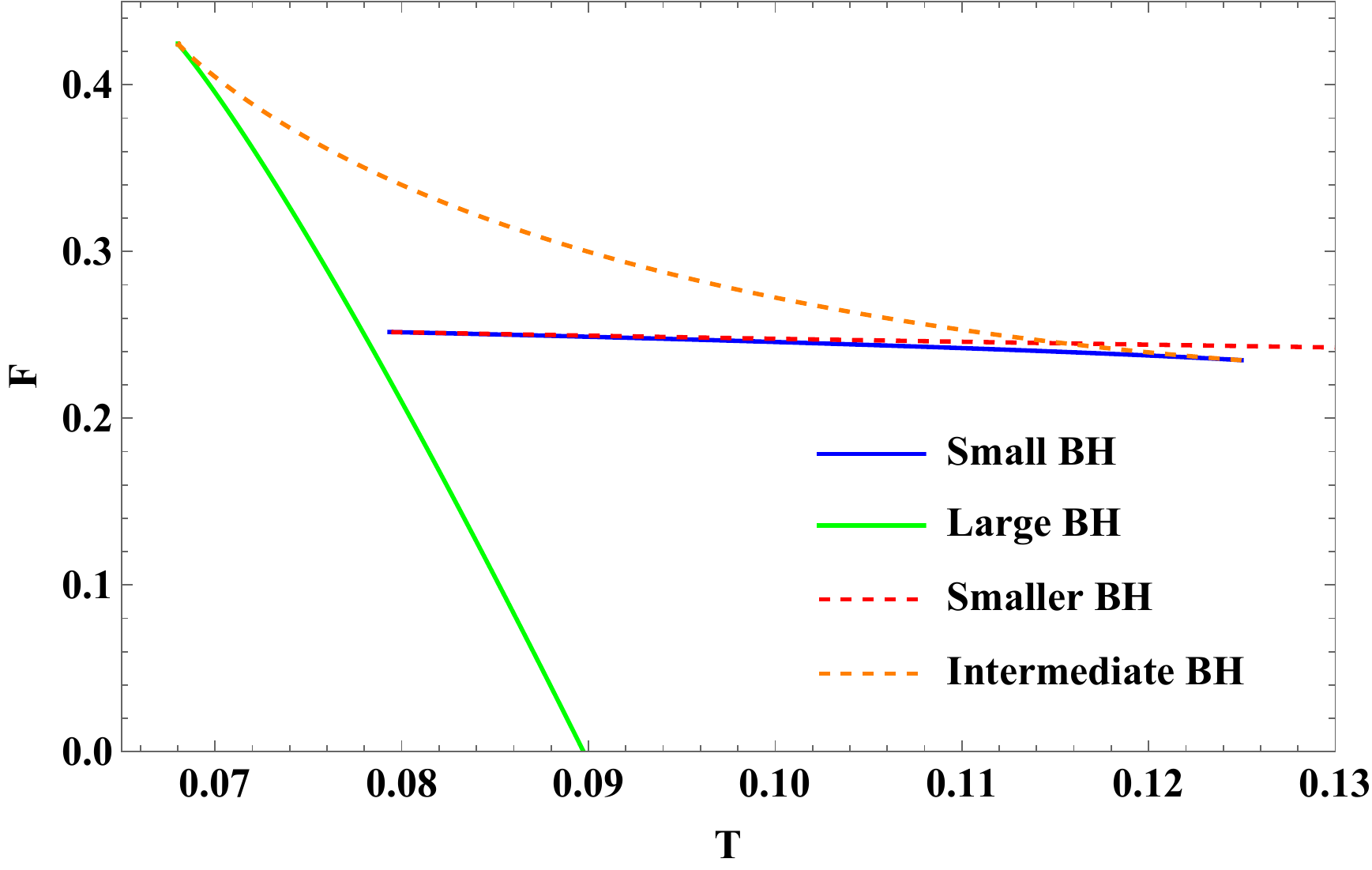}\label{FvTaa}
	\end{minipage}}
	\subfigure[$q_1<q<q_2$]{\begin{minipage}{8cm}
		\includegraphics[width=0.9\linewidth]{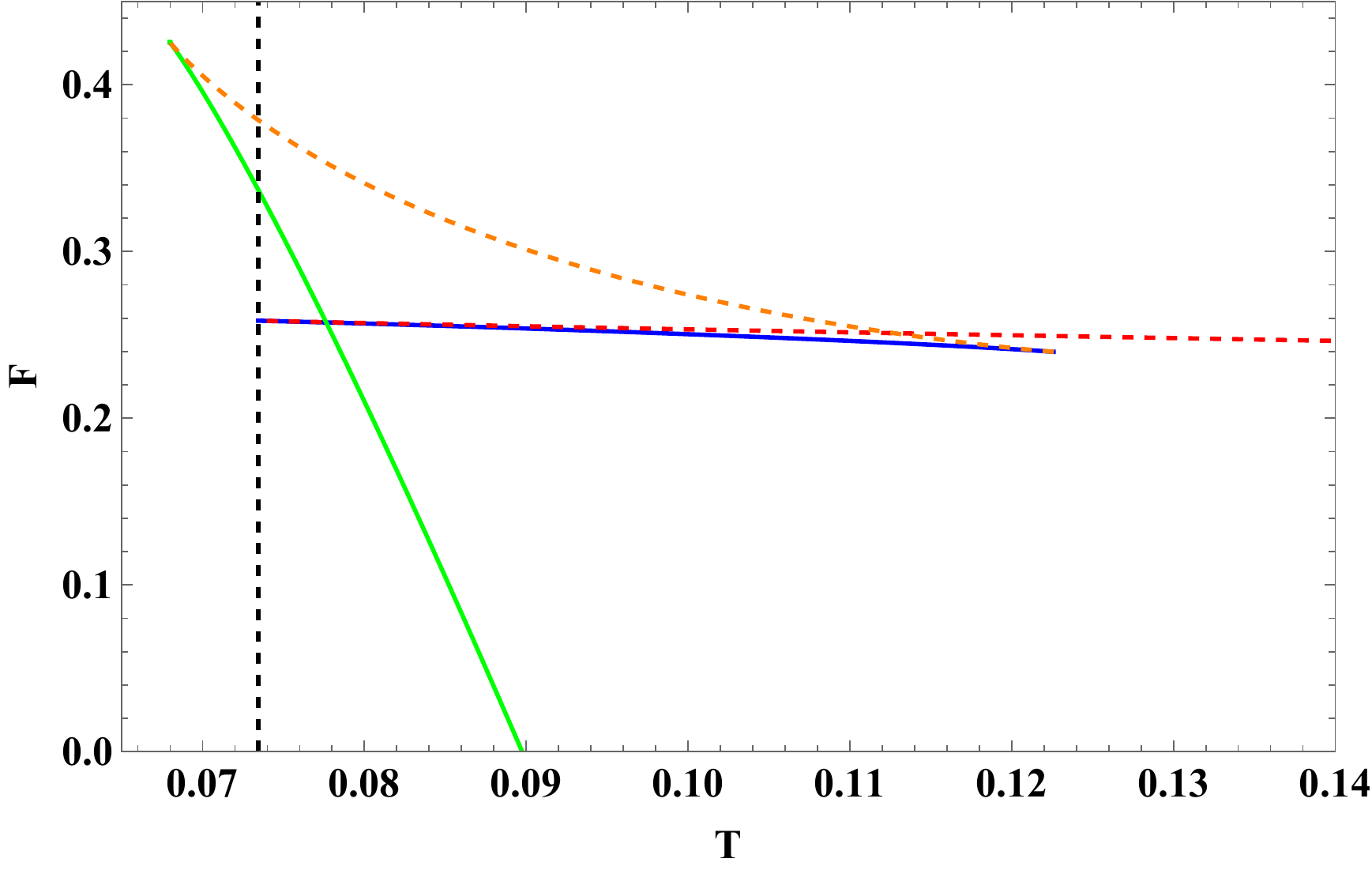}\label{FvTab}
	\end{minipage}}
	\subfigure[$q_2<q<q_{c2}$]{\begin{minipage}{8cm}
			\includegraphics[width=0.9\linewidth]{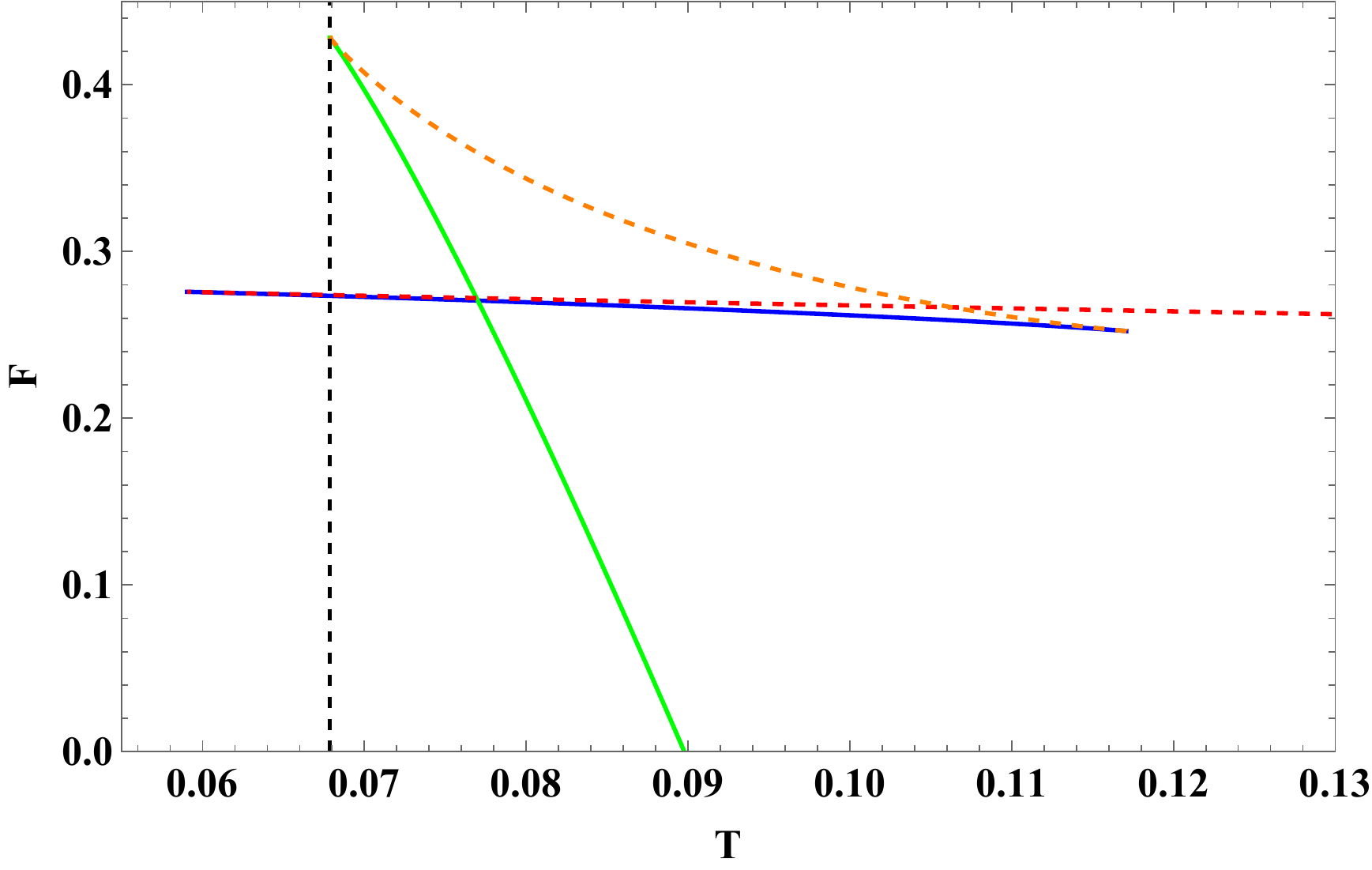}\label{FvTac}
	\end{minipage}}
	\subfigure[$q>q_{c2}$]{\begin{minipage}{8cm}
			\includegraphics[width=0.86\linewidth]{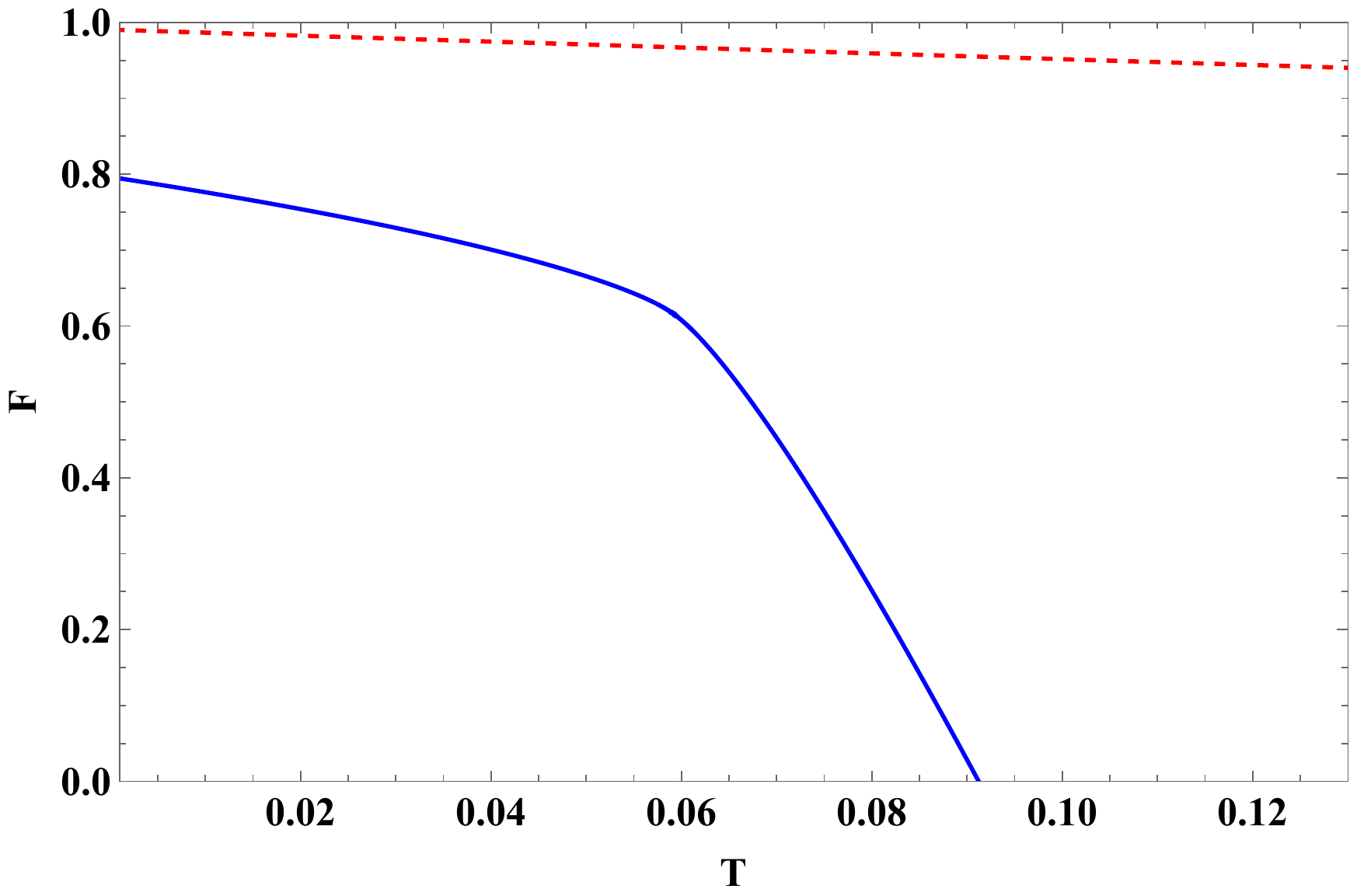}\label{FvTad}
	\end{minipage}}
	\caption{The free energy versus the temperature. \textbf{Upper Left Panel (a):} No phase transition occurs for $q_{c1}<q<q_1$, where we set $q=0.279$. \textbf{Upper Right Panel (b):} The reentrant phase transition for $q_1<q<q_2$, where we set $q=0.285$. \textbf{Lower Left Panel (c):} The first-order phase transition for $q_2<q<q_{c2}$, and we choose $q=0.3$ as an example. \textbf{Lower Right Panel (d):} No phase transition occurs for $q>q_{c2}$, where small and large BH phases can't be distinguished. We set $q=0.8$ here. The dashed curves represent unstable branches, while the solid curves stand for stable or metastable BHs. We set $a=0.08$ and $r_B=3$.}
	\label{FvT-a(0.08)}
\end{figure}

As aforementioned, there will be no phase transition in any case when $a>a_{max}$. For a small positive QED parameter $a<a_{max}$, there are two critical points, as shown in Fig. \ref{Tvrhb}. In this section, we take $a=0.08$ and $r_B=3$, then the value of $q_{c1}$, $q_1$, $q_2$ and $q_{c2}$ can be obtained, which is $0.1839$, $0.2805$, $0.2907$ and $0.7098$, respectively. The phase structures for different charges are depicted in Fig. \ref{FvT-a(0.08)}.

In Fig. \ref{FvTaa}, the analysis of the global minimum free energy shows that there is only one stable BH phase for $q_{c1}<q<q_1$ and no phase transition occurs, which is the same as the case for a small charge.  Here we take $q=0.279$ for example.

The reentrant phase transition, a typical phase behavior including a zeroth-order phase transition and a first-order phase transition, occurs for $q_1<q<q_2$ as shown in Fig. \ref{FvTab}. Here we set the charge $q=0.285$. The solid curves stand for stable or metastable BHs, which can be judged that the former owns the lowest free energy at the same temperature. The dashed curves represent unstable BH branches. The black dashed line corresponds to the temperature of the beginning of stable small BH branch (blue solid curve), which is greater than the initial temperature of the stable large BH branch (green solid curve) in this case. The phase behavior here is a zeroth-order phase transition, where the black hole system jumps from the stable large BH phase to stable small BH phase along the black dashed line. Consequently, it indicates  sudden changes for not only the horizon radius, but also the free energy of the black hole. In addition, by following  the rule of the lowest free energy, the black hole system undergoes a first-order phase transition, namely the small/large BH phase transition. There is only a sudden change in the horizon radius of the black hole as a non-differentiable point in $F-T$ diagram. Analyzing the phase behavior of stable BH branches, we summarize the process of reentrant phase transition in Euler-Heisenberg black holes as follows. The black hole system goes through a zeroth-order phase transition from stable large BH phase to stable small BH phase, and then it returns to stable large BH phase through a first-order phase transition, namely the reentrant large/small/large BH phase transition.

For $q_{2}<q=0.3<q_{c2}$, we find that there is a first-order phase transition from stable small BH phase to stable large BH phase, as shown in Fig. \ref{FvTac}. The other branches with positive heat capacity are metastable phases, corresponding to superheated small BH and supercooled large BH. With the continuous increase of charge, the swallow tail will disappear  and shrink to the critical point $q=q_{c2}$. 

For $q=0.8>q_{c2}$ in Fig. \ref{FvTad}, there is no phase transition, and one cannot distinguish the large BH phase and the small BH phase anymore. 

\begin{figure}[htbp]	
	\centering
	\subfigure[]{\begin{minipage}{8cm}
			\includegraphics[width=0.9\linewidth]{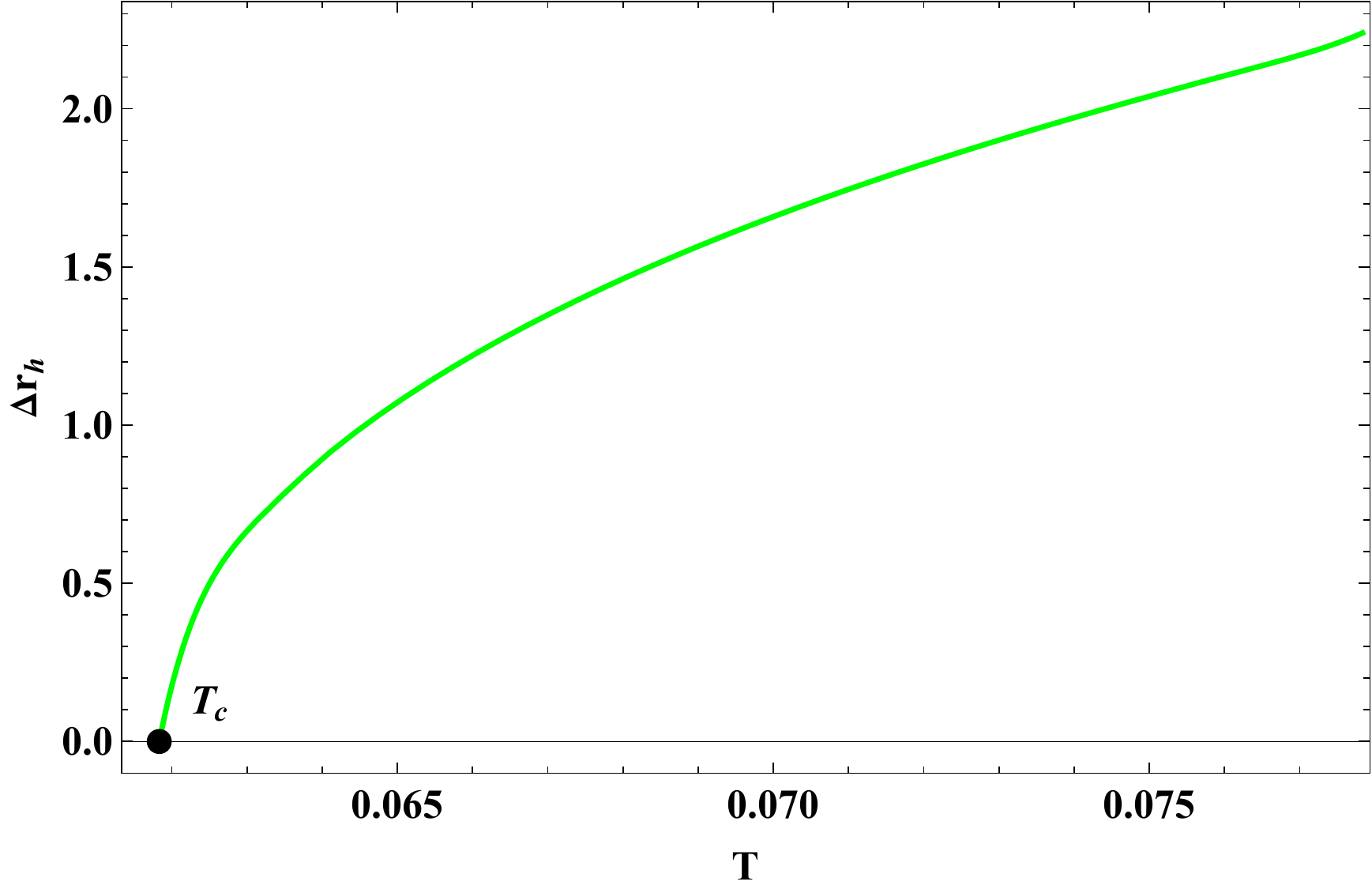}\label{drhvT}
	\end{minipage}}
	\subfigure[]{\begin{minipage}{8cm}
			\includegraphics[width=0.9\linewidth]{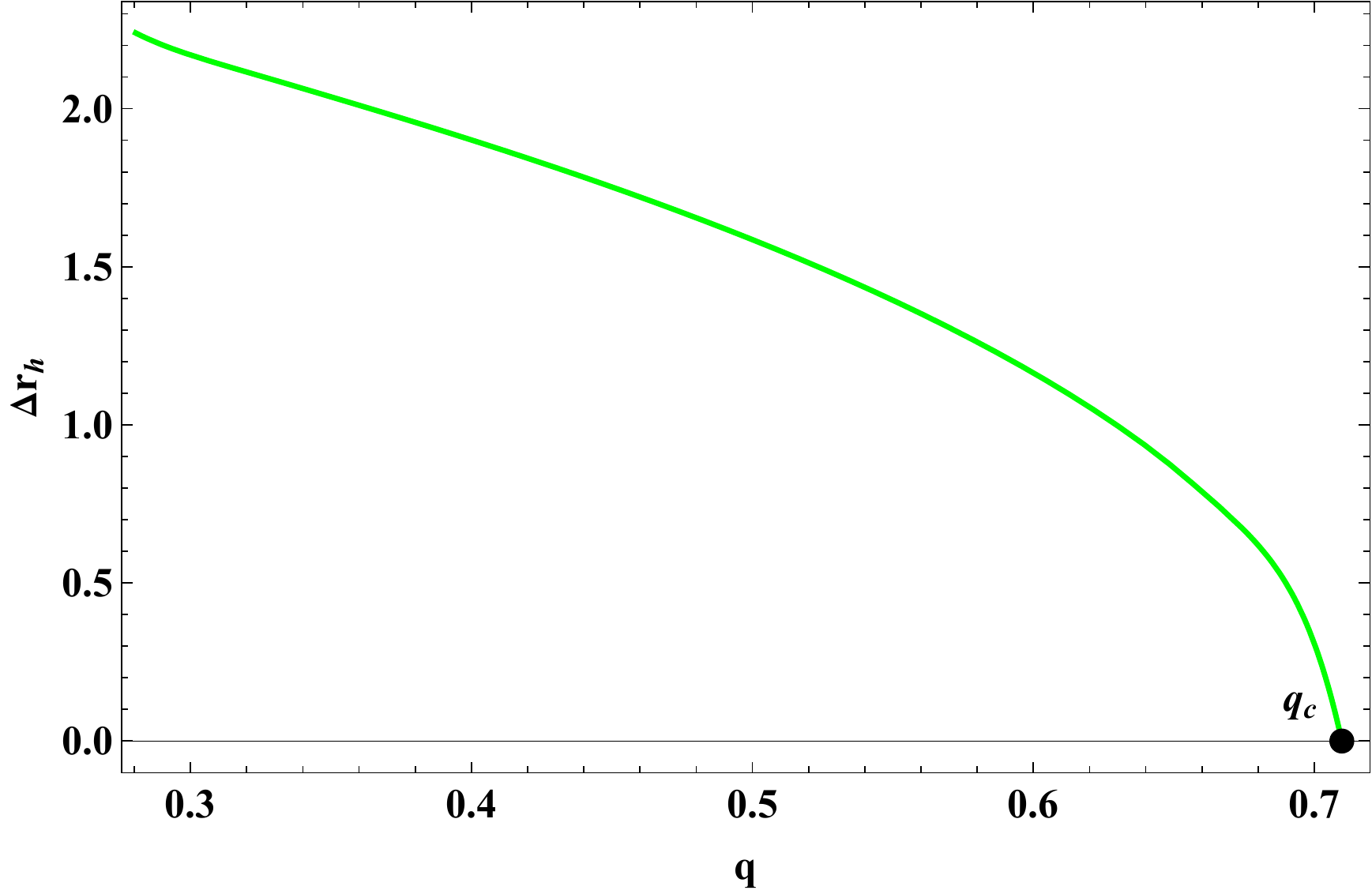}\label{drhvq}
	\end{minipage}}
	\caption{The sudden change $\Delta r_h$ when the black hole comes up a first-order phase transition. \textbf{Left Panel (a):} As the temperature decreases, the $\Delta r_h$ decreases to zero. \textbf{Right Panel (b):} As the charge increases, the $\Delta r_h$ decreases to zero. The QED parameter and radius of the cavity are set to be $a=0.08$ and $r_B=3$.}
	\label{drh}
\end{figure}

To investigate the process from distinguishable BH phase to the indistinguishable one, we take $\Delta r_h=r_{hl}-r_{hs}$ versus the temperature and charge for $a=0.08$ and $r_B=3$ in Fig. \ref{drh}. The parameter $r_{hl}$ is the radius of the event horizon for the large BH branch and $r_{hs}$ is small one. The $\Delta r_h$ represents the sudden change of the horizon radius when the black hole comes up with a first-order phase transition from a small BH phase to a large BH phase. As the temperature decreases or the charge increases, $\Delta r_h$ decreases and eventually turns to zero at the critical point, where the small and large BH phases can not be distinguished any more. 

\begin{figure}[htbp]	
	\centering
	\subfigure[]{\begin{minipage}{8cm}
			\includegraphics[width=0.9\linewidth]{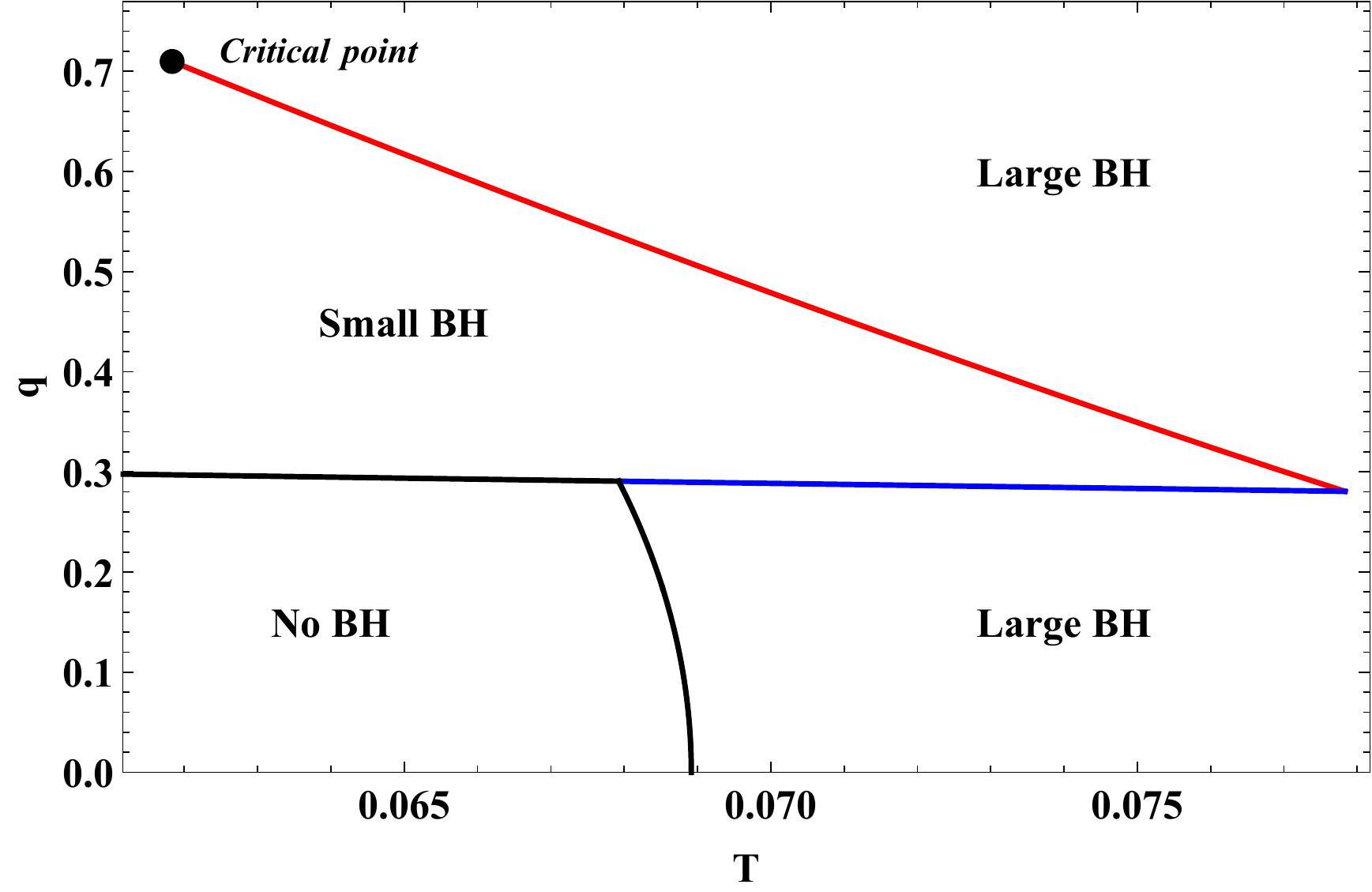}\label{qvTa}
	\end{minipage}}
	\subfigure[]{\begin{minipage}{8cm}
			\includegraphics[width=0.9\linewidth]{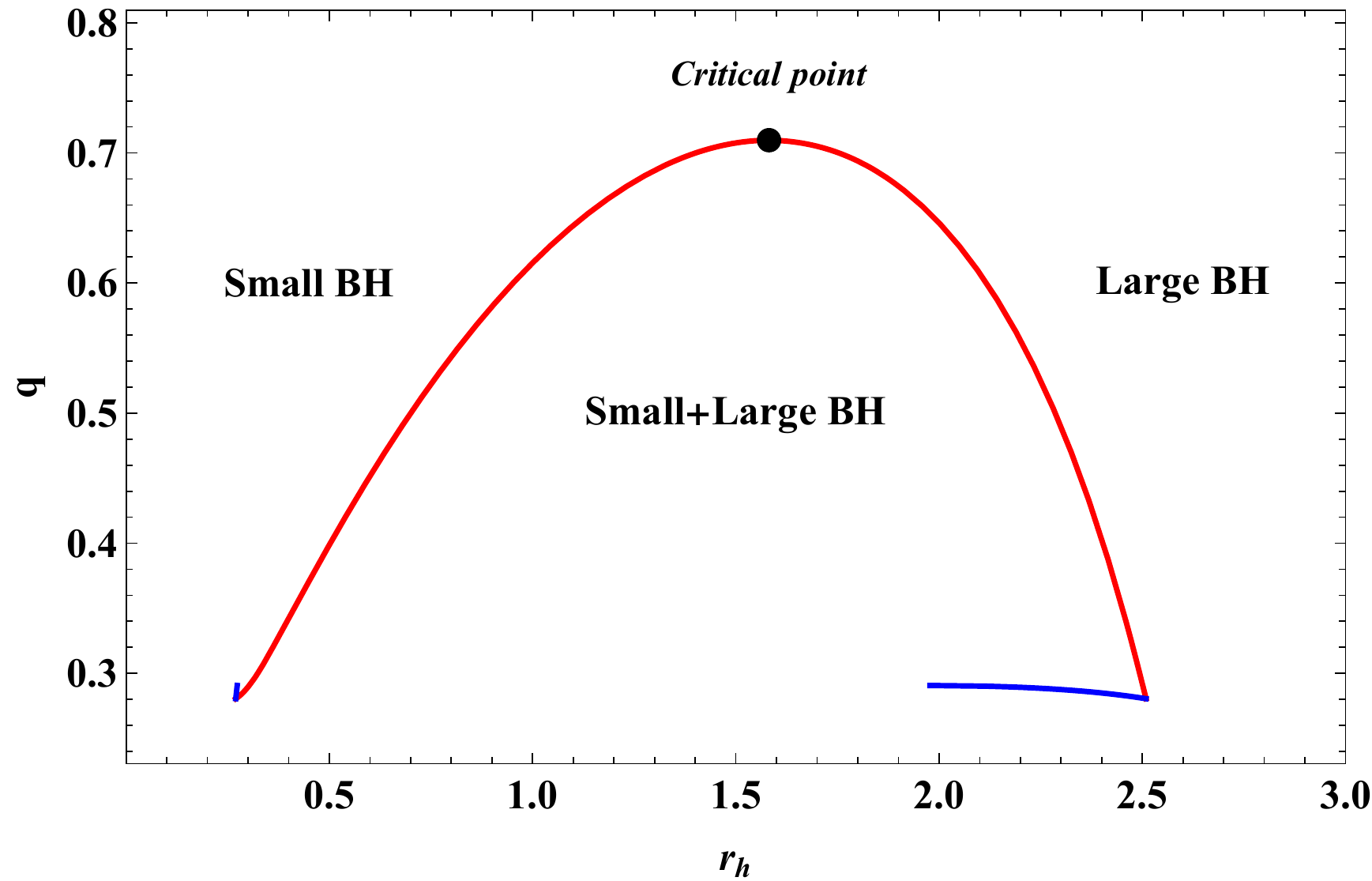}\label{qvrha}
	\end{minipage}}
	\caption{Phase diagram for $a>0$ by taking the charge $q$ as an analogy of the pressure. \textbf{Left Panel (a):} First-order coexistence curve (red solid) and zeroth-order phase transition curve (blue solid) in the $q-T$ plane. The black curve separates the regions with and without black holes. The coexistence curve separates black holes into small and large BH phases, and ends at the critical point of the small/large BH phase transition. \textbf{Right Panel (b):} The phase structure in the $q-r_h$ plane. We choose the parameter $a=0.08$ and the radius of cavity $r_B=3$.}
	\label{qv-a(0.08)}
\end{figure}

The phase structures of Euler-Heisenberg black hole in a cavity are shown in Fig. \ref{qv-a(0.08)}. 
The coexistence curve \cite{Wei2015} is represented by the red solid curve in Fig. \ref{qvTa}. The minimum temperature curve is marked by the black solid curve, which divides regions with or without BH phases. The coexistence curve and the minimum temperature curve are connected by the zeroth-order phase transition curve (blue solid). The small BH region is between the coexistence curve and zeroth-order phase transition curve. The large BH regions are upon the coexistence curve and below the zeroth-order phase transition curve. We show the phase structure in the $q-r_h$ plane in Fig. \ref{qvrha}. The red curve corresponds to the coexistence curve of the small BH phase and large BH phase, while the blue ones denote the zeroth-order phase transition curve. The small and large BH phases are on the left and right sides of the coexistence curve, respectively. The phase coexistence region is bounded by the coexistence curve, which includes the superheated small BH branch, supercooled large BH branch and unstable BH branches. It is noteworthy that below the right branch of zeroth-order phase transition curve, another region for large BH phases exists.

\subsection{$a<0$ case}
\label{sec:3.2}
\begin{figure}[htbp]	
	\centering
	\subfigure[]{\begin{minipage}{8cm}
			\includegraphics[width=0.9\linewidth]{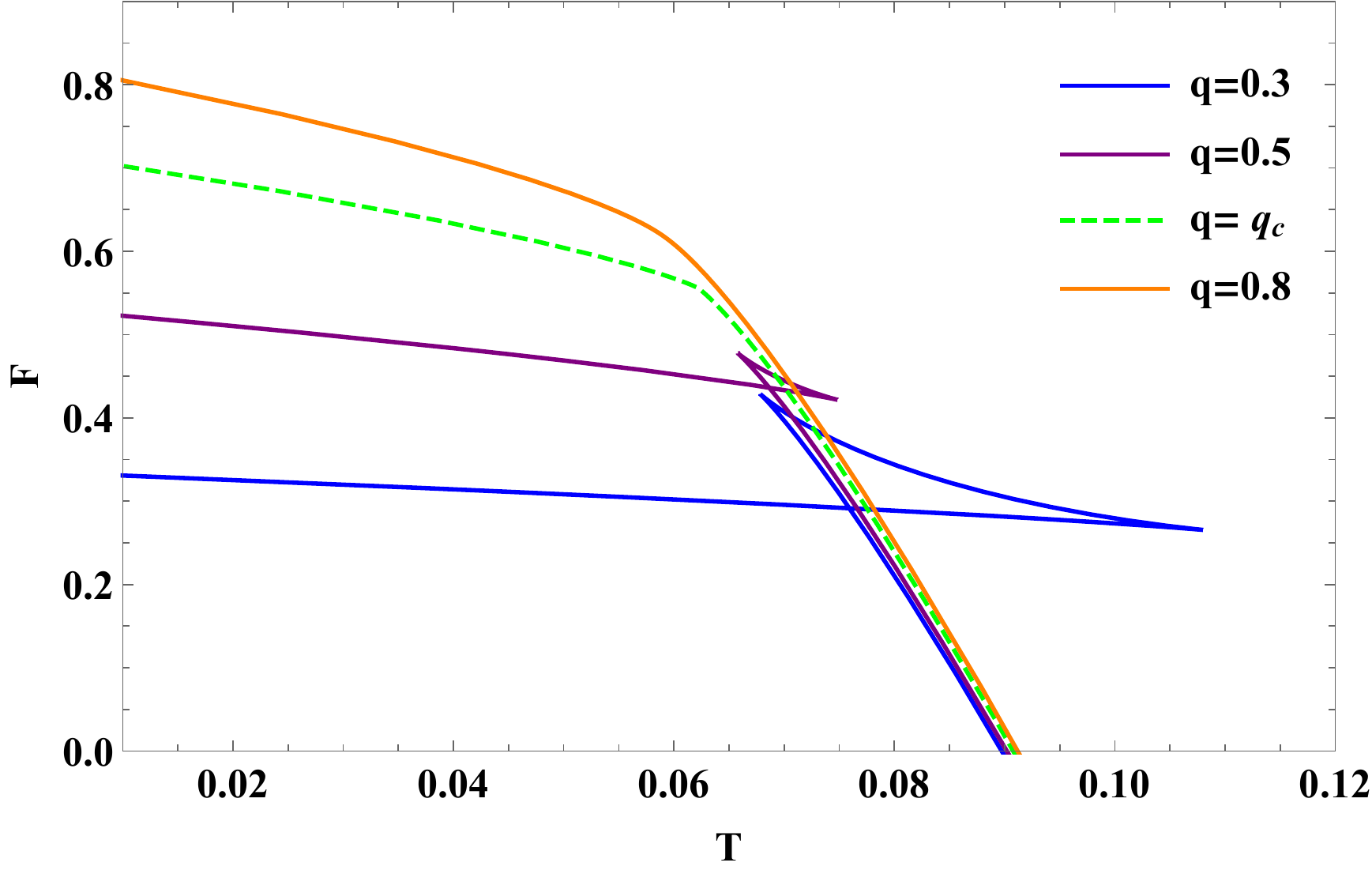}\label{FvTba}
	\end{minipage}}
	\subfigure[$q<q_c$]{\begin{minipage}{8cm}
			\includegraphics[width=0.9\linewidth]{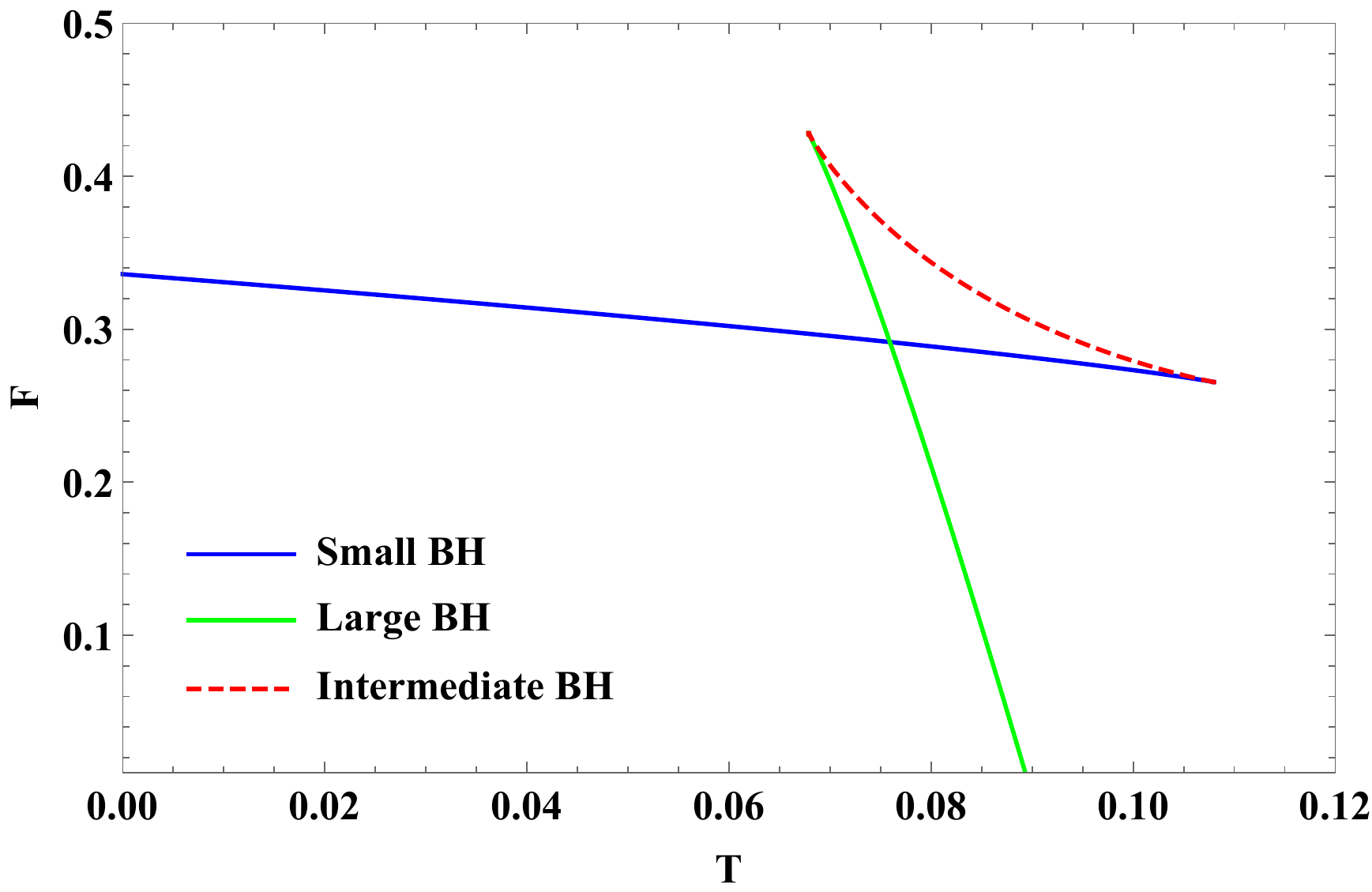}\label{FvTbb}
	\end{minipage}}
	\caption{\textbf{Left Panel (a):} The free energy versus the temperature for $q=0.3$ (blue solid curve), $q=0.5$ (purple solid curve), $q=0.6909$ (green dashed curve) and $q=0.8$ (orange solid curve). \textbf{Right Panel (b):} The free energy versus the temperature for $q<q_c$, where we choose $q=0.3$. The small/large BH phase transition occurs at the intersection of the blue and green solid curve, which represents the stable small BH phase and the stable large BH phase, respectively. The red dashed curve is unstable intermediate BH branch. We set $a=-1$ and $r_B=3$ here.}
	\label{FvT-1}
\end{figure}

The Euler-Heisenberg black hole in a cavity undergoes a single phase transition for a negative QED parameter, which is different from the case for a small positive QED parameter. For the condition that $a=-1$ and $r_B=3$, we can obtain $q_c=0.6909$, and the free energy $F$ versus the temperature $T$ for different values of charge $q$ can be shown in Fig. \ref{FvT-1}. 

The swallow tail behavior occurs when the charge $q<q_c$ in Fig. \ref{FvTba}, indicating that there is a first-order phase transition. With the increase of charge, the swallow tail shrinks to a point at the critical charge $q=q_c$. After that, there will be no phase transition for a greater $q$, where a smooth free energy function $F(T)$ appears. Furthermore, we investigate the free energy versus the temperature for the charge $q=0.3$  in Fig. \ref{FvTbb}. By judging the heat capacity of the black hole, one can tell that the red dashed curve represents the unstable BH phase, while the others are stable ones. As discussed before, the phase transition happens at the intersection of the blue and green solid curves. Since the stable system's free energy is always lower compared to the unstable one, we can conclude that the small/large BH phase transition happens between a small BH phase and a large BH phase.

\begin{figure}[htbp]	
	\centering
	\subfigure[]{\begin{minipage}{8cm}
			\includegraphics[width=0.9\linewidth]{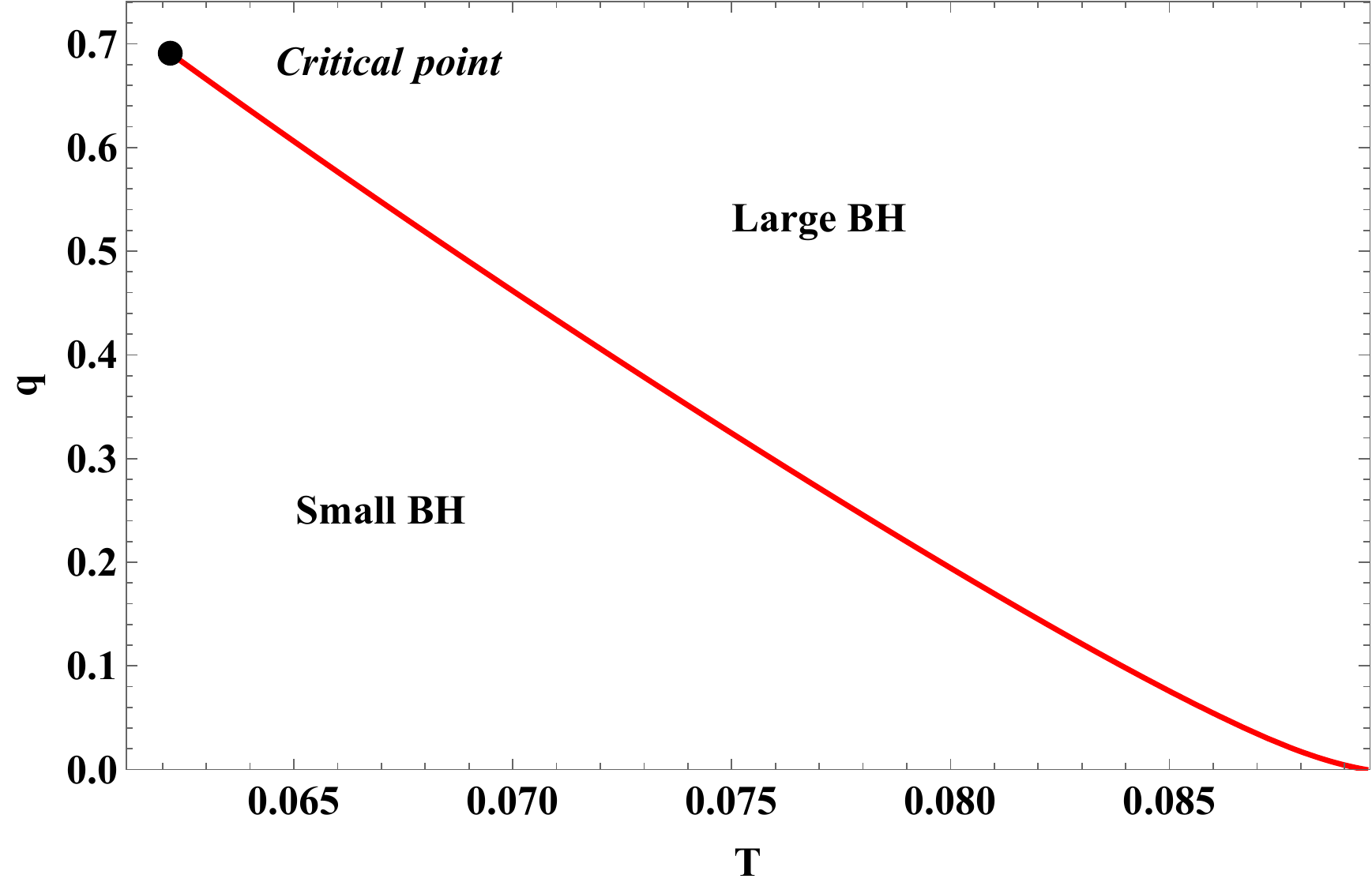}\label{qvTb}
	\end{minipage}}
	\subfigure[]{\begin{minipage}{8cm}
			\includegraphics[width=0.9\linewidth]{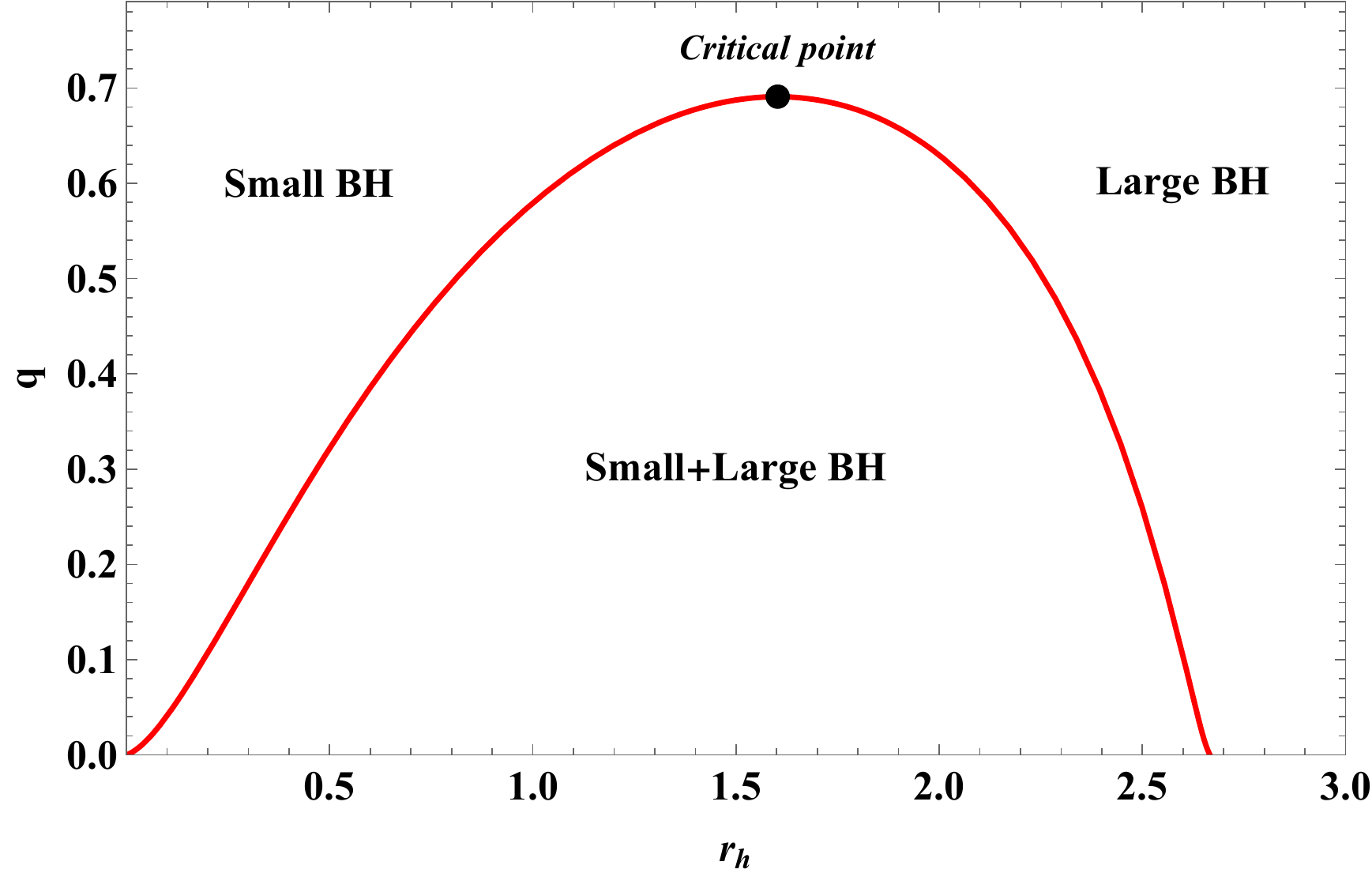}\label{qvrhb}
	\end{minipage}}
	\caption{Phase diagram for $a<0$ by taking charge $q$ as an analogy of pressure. \textbf{Left Panel (a):} First-order coexistence curve in the $q-T$ plane, which separates the black hole into small BH phase and large BH phase. The coexistence curve ends at the critical point of the small/large BH phase transition. \textbf{Right Panel (b):} The phase structure in the $q-r_h$ plane. With the increase of radius, small and large BHs coexist in the region bounded by the red curve and the axis. We choose the parameter $a=-1$ and the cavity radius $r_B=3$.}
	\label{qv-a(-1)}
\end{figure}

The phase structures of Euler-Heisenberg black holes in a cavity for $a<0$ can be shown in Fig. \ref{qv-a(-1)}. The coexistence curve approximately starts at $q=0$, as shown in Fig. \ref{qvTb}, and ends at the critical point where the first-order phase transition disappears. As a coexistence curve, it divides the black hole into small and large BH phases in the $q-T$ phase diagram. In Fig. \ref{qvrhb}, the coexistence region of the small and large BHs occupies below the coexistence curve.

There are some differences between the case that $a<0$ and $0<a<a_{max}$. The swallow tail behavior in $F-T$ diagram shows that the black hole admits a first-order phase transition for $a<0$, such as in Fig. \ref{FvTbb}. This phase transition exists even for a small charge. However, phase transitions do not occur at first for $0<a<a_{max}$ as the charge increases. In addition, the difference in critical points for $a<0$ and $a>0$ may be analyzed from the perspective of topology \cite{Bai2022,Wei2022a}, which is considered as a change of topological charge.

\section{Ruppeiner geometry and microstructure}\label{sec:4}
After analyzing the effect of the QED parameter for phase transitions, we can study the microstructure of the Euler-Heisenberg black hole in a cavity by using the Ruppeiner geometry \cite{Ruppeiner1995}. The Ruppeiner metric $g_{\mu \nu}$ is defined as the negative second derivative of the entropy with respect to the independent variables $x^\mu$ in the thermodynamic system \cite{Ruppeiner1995}, and is given by
\begin{equation}
	g_{\mu \nu}=-\frac{\partial^2 S}{\partial x^\mu \partial x^\nu},
	\label{gab}
\end{equation}
where the $ x^\mu=(U,Y)$ and $U$ is the internal energy and $Y$ is another extensive variable of the system. The Ricci scalar of the Ruppeiner metric is called the Ruppeiner invariant $R$, whose value can shed some light on the microstructure of black holes. In Fig. \ref{Tvrh}, we judge the thermal stability by $C_q$ which is equivalent to the slope of $T(r_h)$, and the Ruppeiner invariant diverges at the divergent point of the specific heat capacity \cite{Mansoori2014}. Hence, we choose the thermodynamic coordinates $x^\mu=(U,\Phi)$ where we define $U$ as a new conjugate potential of $E$. The internal energy is the Legendre transformation of $E$, yielding
\begin{equation}
	U=E-\Phi q.
\end{equation}
The total derivative of $U$ is given by
\begin{equation}
	\mathrm{d}U=T\mathrm{d}S-q\mathrm{d}\Phi.
\end{equation}
By using Eq. (\ref{gab}) and the Maxwell relation $(\frac{\partial T}{\partial q}) {}_S=(\frac{\partial \Phi}{\partial S}) {}_q$, we obtain the line element of the Ruppeiner geometry 
\begin{equation}
	\mathrm{d}s^2=\frac{1}{T}\left(\frac{\partial T}{\partial S}\right)\mathrm{d}S^2-\frac{1}{T}\left(\frac{\partial \Phi}{\partial q}\right)\mathrm{d}q^2.
	\label{line-element}
\end{equation}
Then 
\begin{equation}
	R(S,q)=\frac{A(S,q)+B(S,q)+C(S,q)}{2T (\partial _q \Phi)^2(\partial_S T)^2},
\end{equation}
where
\begin{align}
	A(S,q) & =2\left(\partial_q \Phi \right)\left(\partial_S T\right)^2\left[-\left(\partial_q \Phi \right)\left(\partial_S T\right)+\left(\partial_S \Phi\right)^2\right],\\
	B(S,q) & =T^2\bigg\{\left(\partial_{q,q}T\right)\left(\partial_S T\right)+\left(\partial_q \Phi\right)\left(\partial_{q,q} T\right)\left(\partial_{S,S} T\right)\nonumber\\
	&\quad -\left(\partial_{S,S}\Phi\right)\left[\left(\partial_{q,q}\Phi \right)\left(\partial_S T\right)+\left(\partial_q \Phi \right)\left(\partial_{S,S}\Phi\right)\right]\bigg\},\\
	C(S,q) & =T\left(\partial_S T\right)\bigg\{ \left(\partial_{q,q}\Phi \right)\left(\partial _S T\right)\left(\partial_S \Phi\right)+\left(\partial_q \Phi \right)^2\left(\partial_{S,S} T\right)\nonumber\\
	&\quad -\left(\partial_q \Phi\right)\left[\left(\partial_{q,q}T\right)\left(\partial_S T\right)+\left(\partial_S \Phi\right)\left(\partial_{S,S}\Phi\right)\right] \bigg\} .
\end{align}
The heat capacity is $C_q=T(\frac{\partial S}{\partial T})_q$, so the denominator of $R(S,q)$ can be written as $2T^3(\frac{\partial \Phi}{\partial q})^2C_q^{-2}$. Therefore, singularities of $C_q$ correspond to these of $R$ with $T\neq 0$. The sign of $R$ indicates the dominant interaction between BH molecules : $R>0$ means repulsion (e.g. the ideal fermi gas) while $R<0$ represents attraction (e.g. the ideal bose gas), and $R=0$ implies no interaction (e.g. the ideal gas). The difference between the black hole  phase structure for the positive and negative QED parameters implies that the sign of $a$ affects the microstructure of the black hole. In the following subsections, the behavior of the Ruppeiner invariant for the Euler-Heisenberg black hole in a cavity will be studied for the different sign of $a$.

\subsection{$a>0$ case}
In a certain range of the charge $q$ and positive QED parameter $a$, the system undergoes a reentrant phase transition. As above, we take $a=0.08$ and $r_B=3$ as an example. It has concluded in Sec. \ref{sec:3.1} that only the large BH is the global thermal stable phase for $0<q<q_1$, while a reentrant phase transition occurs for $q_1<q<q_2$, and then a single first-order phase transition replaces in $q_2<q<q_{c2}$, followed by the case in which the small BH and large BH can not be distinguished for $q>q_{c2}$.

\begin{figure}[htbp]
	\subfigure[$q_{c1}<q<q_1$]{\begin{minipage}{8cm}
		\includegraphics[width=1\linewidth]{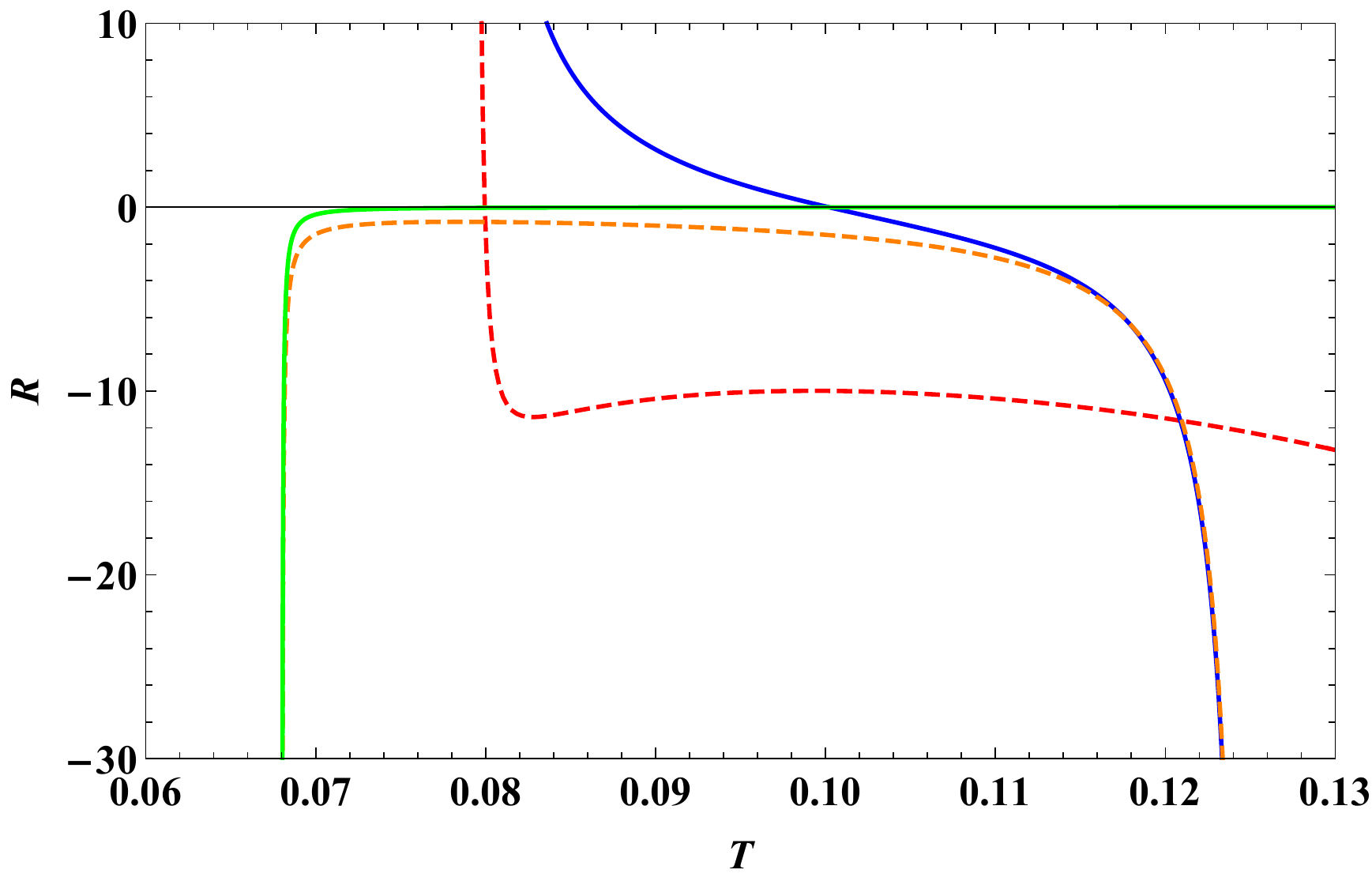}\label{rvt-a(0.08)-q(0.279)}
	\end{minipage}}
	\subfigure[$q_{1}<q<q_{2}$]{\begin{minipage}{8cm}
		\includegraphics[width=1\linewidth]{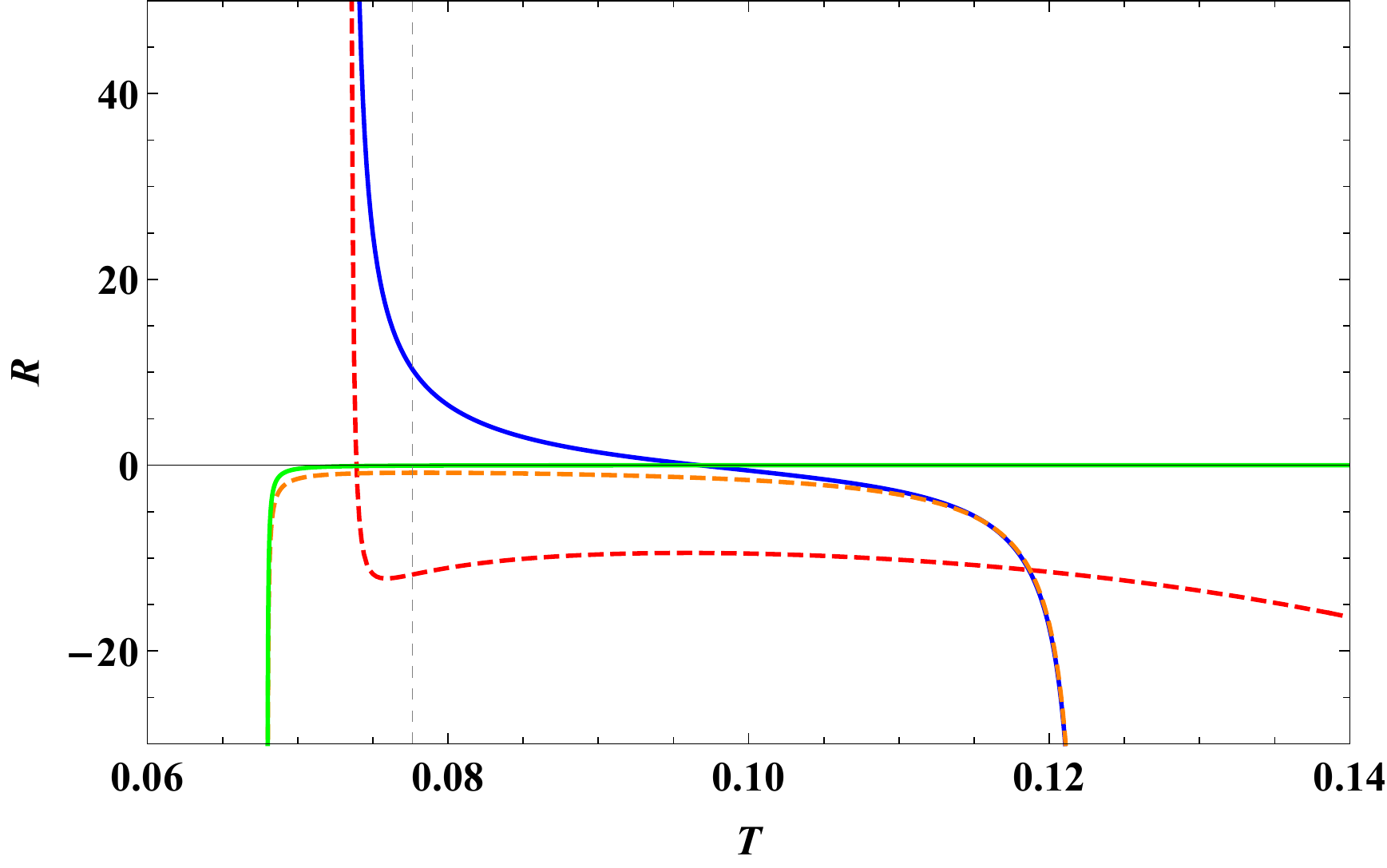}\label{rvt-a(0.08)-q(0.285)}
	\end{minipage}}
	\subfigure[$q_{2}<q<q_{c2}$]{\begin{minipage}{8cm}
		\includegraphics[width=1\linewidth]{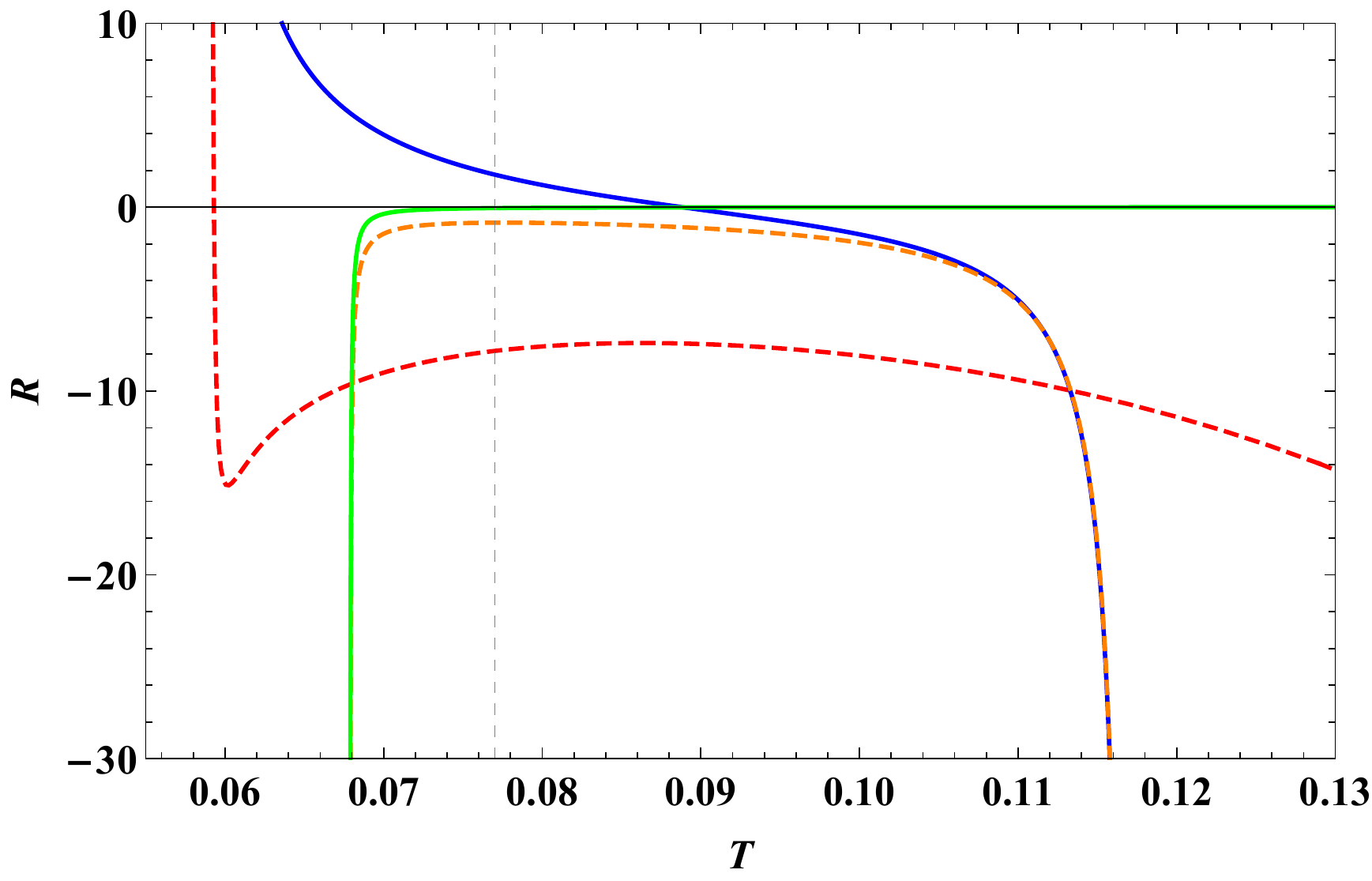}\label{rvt-a(0.08)-q(0.3)}
	\end{minipage}}
	\subfigure[$q>q_{c2}$]{\begin{minipage}{8cm}
		\includegraphics[width=1\linewidth]{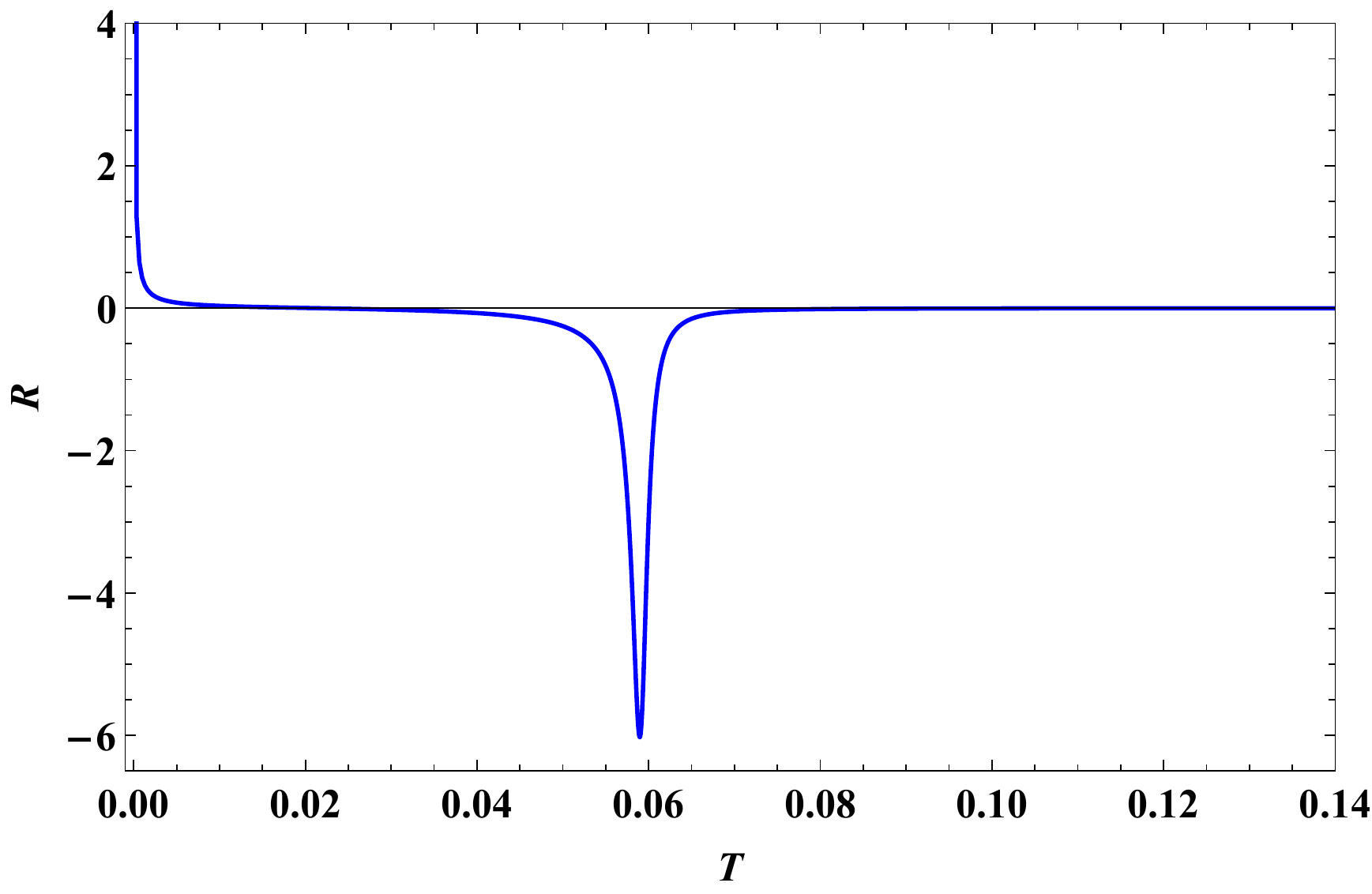}\label{rvt-a(0.08)-q(0.8)}
	\end{minipage}}
	\caption{The Ruppeiner invariant $R$ versus $T$ with a fixed $a=0.08$. \textbf{Upper Left Panel (a):} For $q_{c1}<q<q_{1}$, we set $q=0.279$. \textbf{Upper Right Panel (b):} $q=0.285$, in the range of $q_{1}<q<q_{2}$. \textbf{Lower Left Panel (c):} $q=0.3$ is chosen as an example for $q_{2}<q<q_{c2}$. In these three panels, there are four black hole solutions, and the red/orange dashed curves represent unstable smaller BH and intermediate BH, while the blue/green solid curves represent stable small/large BH, and the black dashed curve corresponds to the temperature of the first-order phase transition. \textbf{Lower Right Panel (d):} $q>q_{c2}$, where we set $q=0.8$, there is only one stable solution (blue solid curve).}
	\label{rvt-a(0.08)}
\end{figure}

We can choose the same values of the charge $q$ as in Sec. \ref{sec:3.1} to show the details of the microstructure of the Euler-Heisenberg black hole in a cavity. According to the phase structure of Fig. \ref{FvT-a(0.08)}, the charge $q=0.279$, $0.285$, $0.3$ and $0.8$ are chosen to present the behavior of the Ruppeiner invariant $R$ in Fig. \ref{rvt-a(0.08)}. 

In the first three panels of Fig. \ref{rvt-a(0.08)}, all of them show that $R$ is always negative for the stable large BH (green solid curve) and the unstable intermediate BH (orange dashed curve), indicating attraction between these BH molecules. But $R$ can be negative or positive for the stable small BH (blue solid curve) and the unstable smaller BH (red dashed curve) depending on the temperature, so the interaction between the BH molecules can be repulsive or attractive.

For $q_{c1}<q<q_1$, the globally stable phase is the large BH with the minimum free energy, so no phase transition happens. In this case as shown in Fig. \ref{rvt-a(0.08)-q(0.279)}, $R$ diverges negatively at the minimum temperature and is always negative for the large BH, demonstrating that the interaction between BH molecules is always attractive for the globally stable phase.

A reentrant phase transition exists for $q_1<q<q_2$, which consists of a first-order phase transition and a zeroth-order phase transition. We can choose $q=0.285$ for this case, and the first-order phase transition between the stable small BH phase and the stable large BH phase occurs for $T=0.0776$, corresponding to the intersection of the small BH and the large BH in Fig. \ref{FvTab}. The Ruppeiner invariant $R$ is positive for the small BH, but it is negative for the large BH when the first-order phase transition happens, as shown in Fig. \ref{rvt-a(0.08)-q(0.285)}. Hence, the type of interaction changes when the system undergoes this phase transition. $|R|$ of the small BH is greater than that of the large BH when the first-order phase transition occurs, so the interaction of the small BH is more intensive than that of the large BH. At the point of the zeroth-order phase transition between the large BH and small BH with $T=0.0734$, corresponding to the black dashed curve in Fig. \ref{FvTab}, $R=+\infty$ for the small BH, while it is negative and finite for the large BH, so it displays that the interaction between BH molecules mutates from weak attraction to intensive repulsion as the temperature increases. 

There is a single first-order phase transition for $q_2<q<q_{c2}$, and the microstructure of this case is shown in Fig. \ref{rvt-a(0.08)-q(0.3)}. The first-order phase transition occurs at $T=0.0770$ for $q=0.3$, corresponding to the intersection of the small BH and large BH in Fig. \ref{FvTac}. $R$ is positive for the small BH but negative for the large BH, and $|R|$ of the small BH is greater than that of the large BH, which indicates that the type of the interaction mutates and attraction of the small BH is more intensive than the repulsion of the large BH at the first-order phase transition point.

No phase transition occurs for $q>q_{c2}$. We could set $q=0.8$ to show the microstructure of the stable BH phase in Fig. \ref{rvt-a(0.08)-q(0.8)}. It shows $R=+\infty$ at $T=0$, and $R\rightarrow 0^-$ with $T\rightarrow +\infty$, which means that the attraction between BH molecules becomes weaker with the increasing temperature generally. The behavior of the Ruppeiner invariant is like the quantum anyon gas \cite{Mirza2009}.

\subsection{$a<0$ case}
\begin{figure}[htbp]
	\subfigure[$q<q_c$]{\begin{minipage}{8cm}
		\includegraphics[width=1\linewidth]{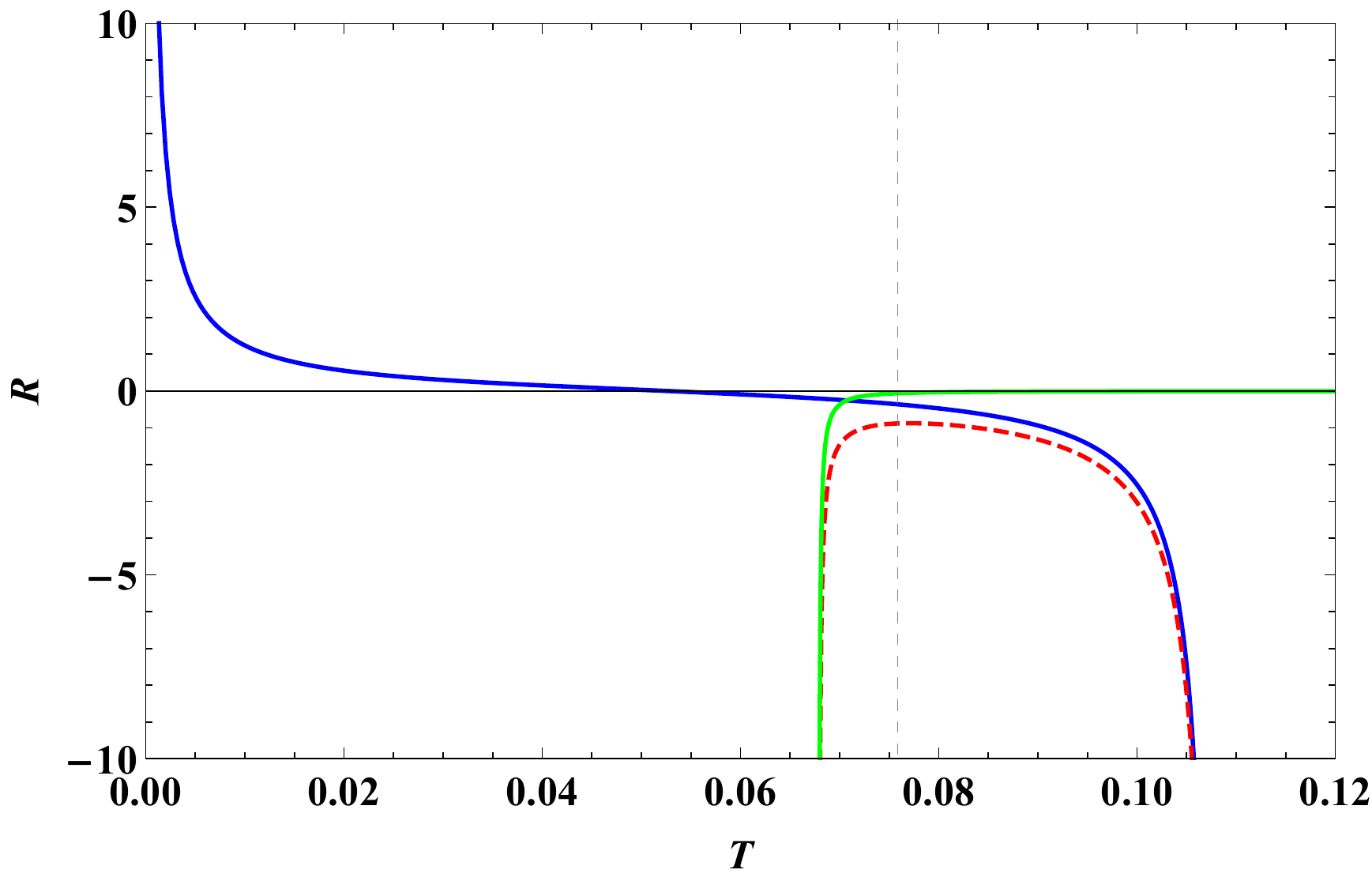}\label{rvt-a(-1)-q(0.3)}
	\end{minipage}}
	\subfigure[$q>q_c$]{\begin{minipage}{8cm}
		\includegraphics[width=1\linewidth]{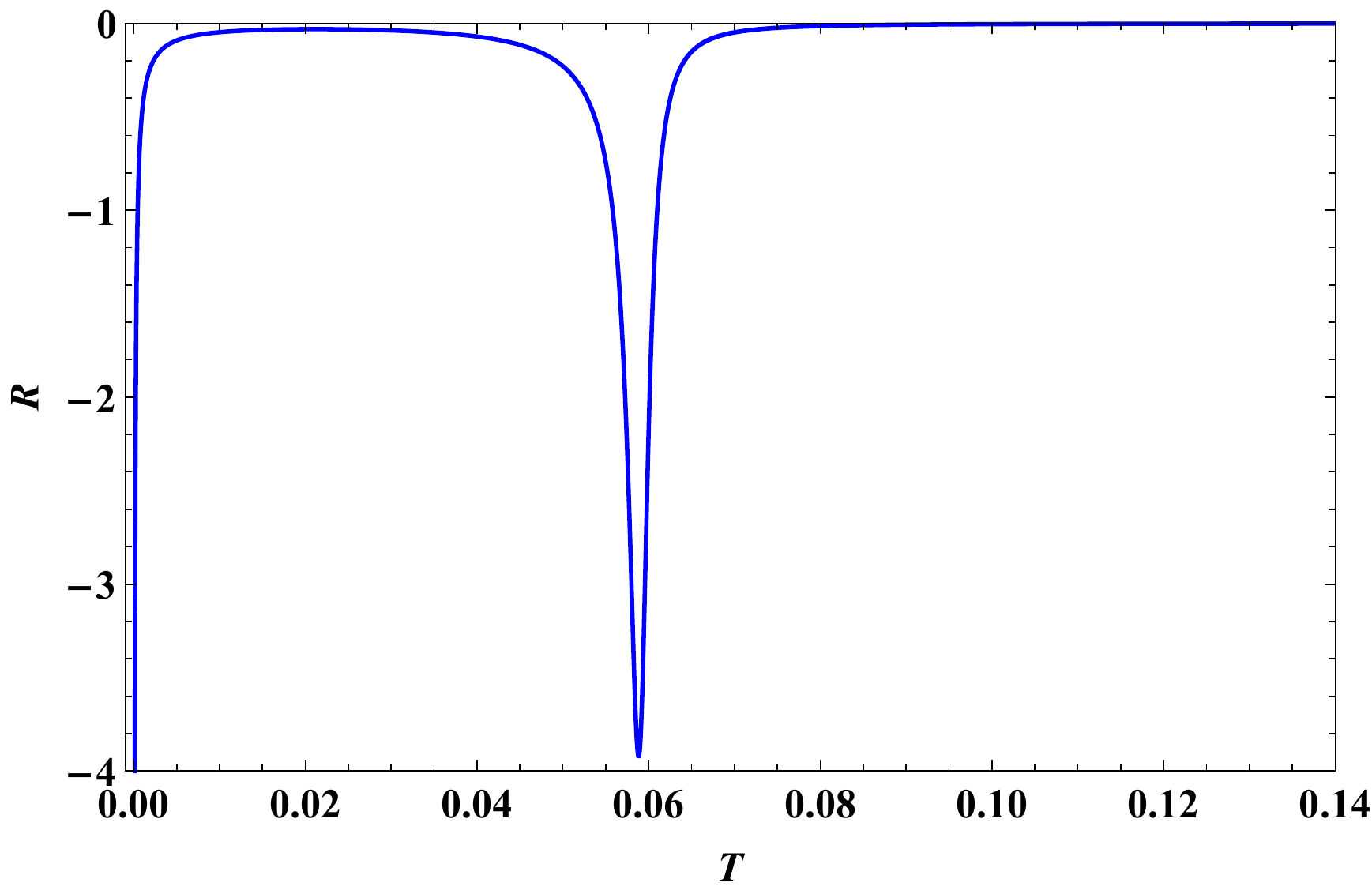}\label{rvt-a(-1)-q(0.8)}
	\end{minipage}}
	\caption{The Ruppeiner invariant $R$ versus temperature $T$ with fixed $a=-1$. \textbf{Left Panel (a):} For $q<q_c$, there are three black hole solutions, and we choose $q=0.3$ for visualization. The red dashed curve represents the unstable intermediate BH, while blue/green solid curve represent the stable small/large BH, and the black dashed curve corresponds to the temperature of the first-order phase transition. \textbf{Right Panel (b):} $q>q_c$, where $q=0.8$ to show the details, there is only one solution and it is stable (blue solid curve).}
	\label{rvt-a(-1)}
\end{figure}

One can explore the microstructure of the black hole systems for $a<0$ by the same method in the last subsection. For the case in Sec. \ref{sec:3.2}, a first-order phase transition exists for a small charge $q$. Choosing $a=-1$ and $r_B=3$ as an example, the system undergoes a first-order phase transition for $0<q<q_c$, while no phase transition occurs for $q>q_c$. For a visual representation, we set $q=0.3$ and $0.8$ to present the behavior of the Ruppeiner invariant $R$ in Fig. \ref{rvt-a(-1)}. The blue and green solid curves represent the stable small BH phase and the stable large BH phase, respectively, and the red dashed curve indicates the behavior of unstable intermediate BH phase. 

For $q<q_c$ as shown in Fig. \ref{rvt-a(-1)-q(0.3)}, $R$ is always negative for the large BH and intermediate BH, so attraction dominates between these BH molecules. Additionally, $R\rightarrow 0^-$ with $T\rightarrow +\infty$ for the large BH indicating the attraction becomes weaker with temperature increasing. However, $R$ can be positive for the small BH with low temperature, so repulsion dominates between these BH molecules. Considering the globally stable phase, one has $R\rightarrow +\infty$ with $T\rightarrow 0$ for the extremal black hole, and $R$ is always finite for $T\neq 0$. A first-order phase transition between the small BH and large BH happens at $T_p=0.0759$, corresponding to the intersection of the small BH and the large BH in Fig. \ref{FvTbb}. At $T=T_p$, $R$ is negative for both the small BH and the large BH, but $|R|$ of the small BH is greater than that of the large BH. Thus, the attraction of the small BH is more intensive than that of the large BH.

For $q>q_c$, the microstructure of the black hole systems is displayed in Fig. \ref{rvt-a(-1)-q(0.8)}. $R\rightarrow +\infty$ with $T\rightarrow 0^+$ , and $R\rightarrow 0^-$ with $T\rightarrow +\infty$. The interaction of the BH molecules is always attractive, like the bose gas. Hence, in the case of no phase transition, the microstructure of the Euler-Heisenberg black hole in a cavity for $a<0$ is different from that of RN black hole in a cavity studied in Ref. \cite{Wang2020}, in which $R$ is always finite in a globally stable phase. 

We can compare the microstructure of the small BH with that of the large BH when the first-order phase transition occurs between $a>0$ and $a<0$. In the case of $a=0.08$, as the first-order phase transition occurs, the small BH intermolecular interaction is repulsive. However, in the case of $a=-1$, attraction dominates the small BH when the first-order phase transition occurs. According to these differences, the Ruppeiner invariant along the first-order coexistence curve is worth studying.

\begin{figure}[h]
	\subfigure[$a>0$]{\begin{minipage}{8cm}
		\includegraphics[width=1\linewidth]{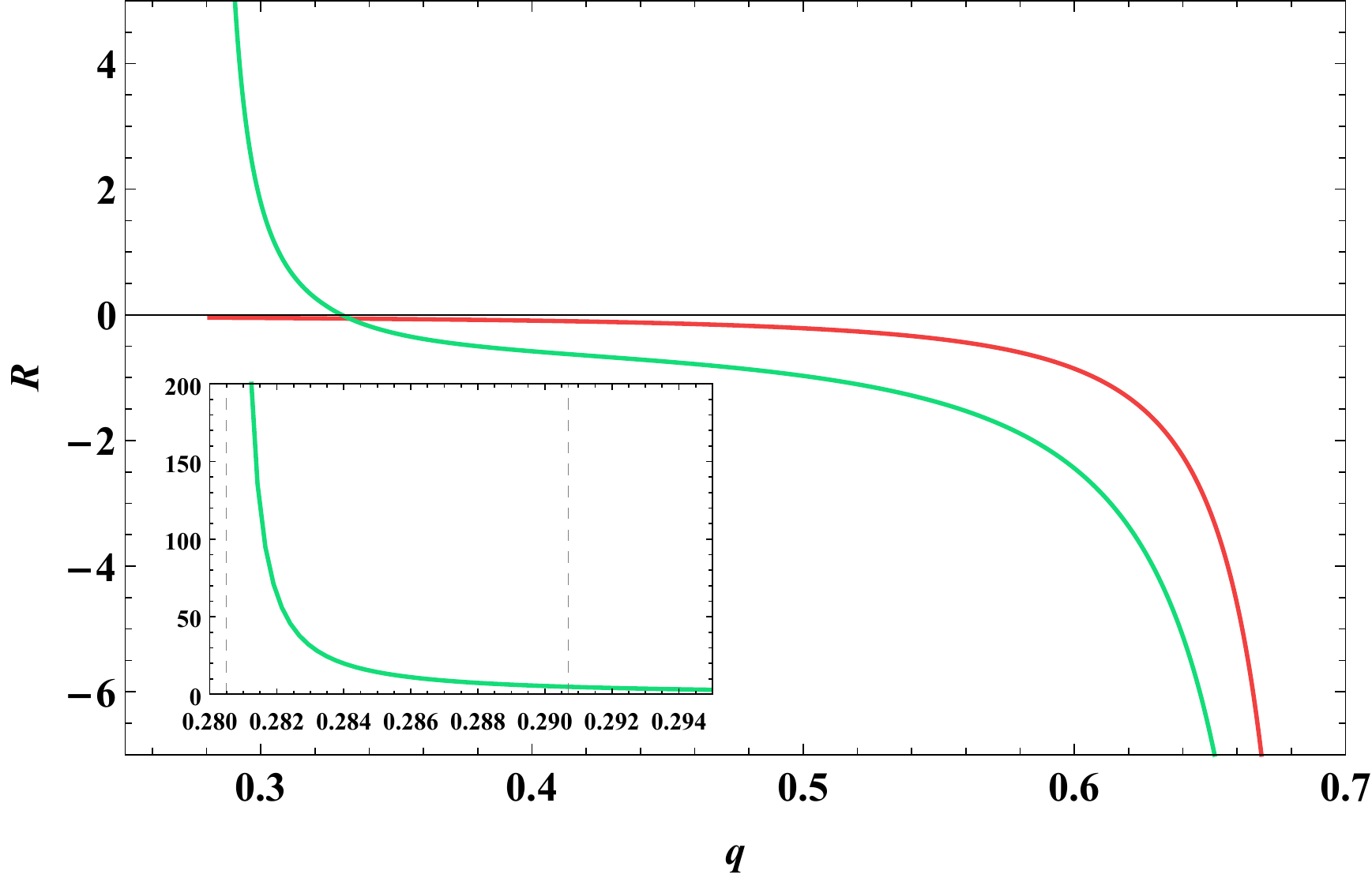}\label{rvq-a(0.08)}
	\end{minipage}}
	\subfigure[$a<0$]{\begin{minipage}{8cm}
		\includegraphics[width=1\linewidth]{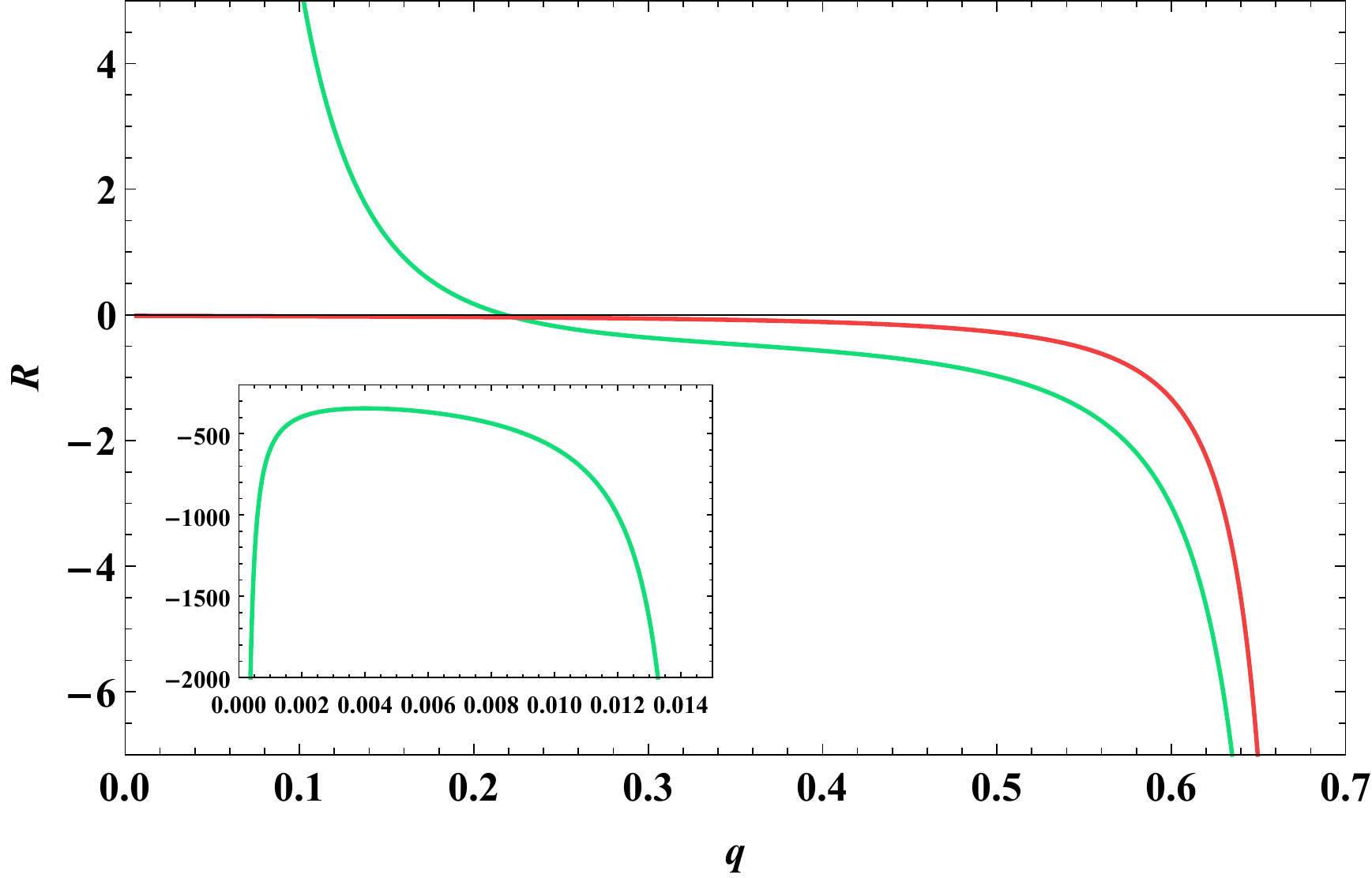}\label{rvq-a(-1)}
	\end{minipage}}
	\caption{The behavior of the Ruppeiner invariant $R$ along the first-order coexistence curve. \textbf{Left Panel (a):} $a=0.08$, $R$ depends on $q$ along the first-order coexistence curve in Fig. \ref{qvTa}. \textbf{Right Panel (b):} $a=-1$, the first-order coexistence curve is in Fig. \ref{qvTb}. The green curve represents the small BH, while the red curve represents the large BH.}
	\label{rvq}
\end{figure}

Then, to study the effect of QED parameter on the microscopic structure more deeply, we examine the behavior of the Ruppeiner invariant along the first-order coexistence curve, as illustrated in Fig. \ref{rvq}. For both the cases of $a>0$ and $a<0$, the interaction of the large BH (the red curve) is always attraction when the first-order phase transition occurs, while the interaction of the small BH (the green curve) is able to be repulsive with the small charge, i.e. the high temperature. In these two cases, the Ruppeiner invariant $R$ diverges negatively at the critical point for both the small BH and large BH, and this feature is similar to the behavior of the Ruppeiner invariant for the Euler-Heisenberg-AdS black hole \cite{Ye2022}. Along the first-order coexistence curve, the microscopic behaviors for the large BH phase of $a>0$ and $a<0$ cases closely resemble, but there are some differences for the small BH phase.

As mentioned before, a reentrant phase transition exists for $q_1<q<q_2$, and the special range is shown in the inset of Fig. \ref{rvq-a(0.08)}. It is obvious that repulsion dominates in this specific range as the reentrant phase transition occurs, and $R$ diverges positively with $q=q_1$. As the charge increases, only the first-order transition exists, while the dominant interaction becomes attractive from repulsive. A change from low to high correlation (the magnitude of $R$) among components is one of the signs of a phase transition (or vice versa) \cite{KordZangeneh2018}. For a given charge, decreasing the temperature, the phase transits from the large BH to the small BH in the most range of $q$, so the $|R|$ becomes greater and the correlation increases. Additionally, in the range of $q_1<q<q_2$, the difference of $|R|$ between the small BH and the large BH is greater than other ranges of charge, as shown in Fig. \ref{rvq-a(0.08)}. Hence, we deduce that the greater difference in the correlation makes a reentrant phase transition occur in the system, which is the same as the conclusion in Ref. \cite{KordZangeneh2018}.

Only a first-order phase transition occurs in the case of $a<0$. The divergent points of the Ruppeiner invariant are three instead of two for the small BH along the first-order coexistence curve. The divergent points correspond to $q=0$, $q_d$ and $q_c$, where $q_d=0.0145$. In the range of $q_d<q<q_c$, the microscopic behavior of this case is similar to that in the case of $a>0$. The inset of Fig. \ref{rvq-a(-1)} shows that the interaction between the small BH molecules is intensively attractive with $0<q<q_d$. 

Therefore, for the reentrant phase transition, only repulsion exists between the BH molecules, but both repulsion and attraction can dominate with the presence of the usual small/large BH phase transition. And the distinction between the cases of negative and positive QED parameters is the extra range dominated by the attraction in the case of $a<0$, as illustrated in the inset of Fig. \ref{rvq-a(-1)}. So we deduce that the attraction may prevent the BH from undergoing an extra zeroth-order phase transition when the temperature decreases, and the negative QED parameter may help the formation of the extra attraction region to obstruct the reentrant phase transition.

\section{Conclusion}
\label{sec:Discussion-and-Conclusions}
In this paper, we focus on the Euler-Heisenberg black hole with the QED correction in a cavity to explore the thermodynamic properties. The phase transitions were investigated for different charges and QED parameters, and the Ruppeiner geometry was analyzed according to the phase structures.

We derived the first law of black hole thermodynamics via the thermal energy, and the free energy of black holes was given naturally. According to the free energy, the phase transitions of Euler-Heisenberg black holes in a cavity were analyzed, which was related to different charges $q$ and QED parameters $a$. Based on this analysis, we found that there is a maximum value of $a$ allowing for phase transitions, which is determined by the cavity radius. And we investigated the temperature and critical points of the black hole systems for $a>0$ and $a<0$, respectively. Then the phase transitions were investigated in detail. For a small positive QED parameter, there could be two critical points and four BH branches, including two stable branches and two unstable branches. As the charge increases, it can be described as follows. The black hole system for a fixed QED parameter undergoes a process from a stable BH phase to a reentrant phase transition, to a first-order phase transition, and thereafter no phase transition occurs. For a negative QED parameter, there is a small/large BH phase transition, and only one critical point exists. This first-order phase transition occurs for a small charge, where there are two stable BH branches and one unstable intermediate BH branch. There is no phase transition for the charge above the critical point.

Further more, we explore the microstructure of this black hole by the Ruppeiner approach \cite{Ruppeiner1995}. As a preliminary step, taking $(U,\Phi)$ as the coordinates in the thermodynamic phase space and defining a conjugate potential as the internal energy, we calculate the Ricci scalar of the Ruppeiner geometry for the black hole. Along the first-order coexistence curve, for the small BH, only repulsion dominates between the BH molecules when the reentrant phase transition occurs, while the dominant interaction can be attractive or repulsive when the usual small/large BH phase transition exists. Then an attraction region in the small BH phase is missing for the positive QED parameter. It infers that the attraction may affect the reentrant phase transition, and the negative QED parameter makes the extra attraction region for the small BH phase, so the zeroth-order phase transition is obstructed by the attraction region.

Following the investigation, it is inspiring to find that the phase structure of the Euler-Heisenberg black hole in a cavity is similar to the black hole with the AdS boundary \cite{Ye2022}. Then, one can note that different thermodynamic coordinates may result in various microstructures, such as the investigation of the Kerr-Newman black hole in Ref. \cite{Huang2022}. So we will focus on the correlation between the phase structures and various microstructures in our future works.

\begin{acknowledgments}
We are grateful to Yihe Cao, Yuchen Huang, Ningchen Bai, Aoyun He and Peng Wang for useful discussions and valuable comments. This work is supported by NSFC (Grant No.12047573 and 12275184).
\end{acknowledgments}

\bibliographystyle{unsrturl}
\nocite{*}
\normalem
\bibliography{eh}

\end{document}